\documentclass[letterpaper,journal]{IEEEtran}
\usepackage{amsmath,amsfonts,amssymb}
\usepackage{algorithmic}
\usepackage{algorithm}
\usepackage{array}
\usepackage{textcomp}
\usepackage{stfloats}
\usepackage{url}
\usepackage{graphicx}
\usepackage{cite}
\usepackage{color}
\usepackage{bm}
\usepackage[font=normalsize,labelfont=bf,labelsep=colon]{caption}
\usepackage[font=normalsize,labelfont=normalfont,textfont=normalfont]{subfig}
\begin{document}
\bstctlcite{BSTcontrol}

\title{\fontsize{23pt}{27pt}\selectfont FreqTune-PASS: Frequency-Tuned Beamforming for Multi-User Pinching-Antenna Systems}

\author{
Minghao Jin, Anna Li,~\IEEEmembership{Member,~IEEE}, Tianwei Hou,~\IEEEmembership{Member,~IEEE},\\ Qiang Ni,~\IEEEmembership{Senior Member,~IEEE}, Mugen Peng,~\IEEEmembership{Fellow,~IEEE}

\thanks{Minghao Jin, Anna Li and Qiang Ni are with the School of Computing and Communications, Lancaster University, Lancaster LA1 4WA, U.K. (e-mail: m.jin6@lancaster.ac.uk; a.li16@lancaster.ac.uk; q.ni@lancaster.ac.uk).}
\thanks{Tianwei Hou and Mugen Peng are with the Beijing Key Laboratory of Convergent Communications and Networking Technologies in LEO Satellite Systems, and also with the School of Information and Communication Engineering, Beijing University of Posts and Telecommunications, Beijing, 100876, China. (e-mail: htw@bupt.edu.cn; pmg@bupt.edu.cn).}
}

\maketitle

\begin{abstract}
Pinching-antenna systems (PASS) reconfigure large-scale propagation environments by activating low-cost radiation points along a dielectric waveguide. However, existing multi-user PASS designs rely on pinching-antenna (PA) position optimization or activation, whose mechanical reconfiguration is much slower than user switching in time-division multiple access systems. To overcome this mismatch, a frequency-tuned beamforming framework (FreqTune-PASS) is proposed, where all PA positions remain fixed and only a user-specific carrier-frequency offset is adjusted in each slot. Since each PA-radiated signal experiences a distinct combination of in-waveguide and free-space delays, a common frequency offset induces path-specific phase rotations and thus reshapes the coherent superposition at the scheduled user. Based on a delay-domain system model, the fast-timescale frequency-offset optimization and the slow-timescale PA-deployment problems are formulated. A closed-form globally optimal offset is derived for the two-PA case, while a low-complexity one-dimensional search and an approximate closed-form offset under equal-spacing deployment are developed for the multi-PA case. The deployment analysis further shows that full-aperture equal spacing maximizes the guaranteed phase-tuning span and yields near-arithmetic relative delays for uniformly distributed users. Numerical results demonstrate that the proposed framework substantially improves both the sum rate and the worst-user rate over fixed-frequency transmission and approaches the coherent-combining upper bound.
\end{abstract}


\section{Introduction}

\IEEEPARstart{T}{he} sixth-generation (6G) wireless networks are expected to provide high spectral efficiency, ubiquitous coverage, and reliable connectivity for densely distributed users \cite{Tataria6GProcIEEE2021}, and these requirements are further escalated by emerging intelligence-native services such as generative artificial intelligence (GAI)- and large language model (LLM)-enabled communications \cite{ChenGAISemanticVideo2025, ChenSecureMLLM2025, ChenLLMSemantic2025}. 
Over the past decades, massive multiple-input multiple-output (MIMO) and millimeter-wave (mmWave) technologies have substantially enhanced network capacity by exploiting large antenna apertures and abundant spectrum \cite{MarzettaMassiveMIMO2010, HeathMmWaveMIMO2016}. Nevertheless, conventional arrays are geometrically fixed once deployed, and the propagation environment is essentially treated as an uncontrollable medium to which the transceiver can only adapt \cite{DiRenzoSmartRadio2020}.

To introduce reconfigurability into the wireless channel, several flexible-antenna paradigms have been developed. Reconfigurable intelligent surfaces (RISs) reshape the propagation environment through nearly passive reflection \cite{LiuRISSurvey2021}, while holographic MIMO and continuous-aperture arrays (CAPAs) push the antenna architecture toward spatially continuous electromagnetic processing \cite{SanguinettiWDMHMIMO2023, HouCAPA2026}. More recently, fluid antenna systems (FASs) \cite{NewFASTutorial2025} and movable antennas (MAs) \cite{ZhuMovableTutorial2026} have introduced position flexibility by selecting or relocating the radiating elements within a confined local region, thereby exploiting spatial degrees of freedom unavailable to fixed arrays.
However, confined to wavelength-scale apertures or fixed installation sites, these technologies can hardly reshape the large-scale propagation conditions, e.g., re-establishing a direct line-of-sight (LoS) link for a user blocked by moving obstacles.
Pinching-antenna systems (PASS) overcome this limitation by operating at a fundamentally different spatial scale \cite{LiuPASSTutorial2026}.
First prototyped by NTT DOCOMO, a pinching antenna (PA) is realized by attaching a small dielectric particle to a dielectric waveguide, thereby creating a low-cost radiation point at the pinched position \cite{FukudaPinchingAntenna2022}. 
Since the waveguide can stretch across the entire ceiling of a room, PAs can be flexibly activated at different positions along it, such that the large-scale path loss becomes a reconfigurable parameter and strong LoS links can be created in blockage-sensitive high-frequency bands without additional radio-frequency chains \cite{DingFlexiblePASS2025}.

These advantages have stimulated extensive research on PASS. For the single-user case, the achievable rate, outage probability, and array gain have been analyzed in \cite{DingFlexiblePASS2025, TyrovolasPASSPerformance2026, OuyangArrayGainPASS2025}, a physics-based hardware model with joint transmit and pinching beamforming was developed in \cite{WangPASSModeling2025}, and PA placement optimization, learning-based beamforming, and comparisons with RIS-assisted mmWave systems were studied in \cite{XuRateMaxPASS2025, XiePlacementPASS2025, GuoPASSGNN2025, SamyPASSRIS2025}. 
For multi-user transmission, existing works cover PA activation for downlink non-orthogonal multiple access (NOMA) \cite{WangNOMAAntennaActivation2025}, uplink PA-position and resource optimization together with performance analysis \cite{TegosMinRatePASS2025,HouUplinkPASS2026}, capacity characterization of single-waveguide PASS \cite{OuyangCapacityPASS2026}, multi-user transmission structures with joint transmit and pinching beamforming \cite{ZhaoPASSMultiuser2026}, waveguide-division multiple access (WDMA) and multi-waveguide antenna activation and resource allocation \cite{ZhaoWDMAPASS2026,WangMultiWaveguidePASS2026,ZhangWDMAorNOMA2026}, as well as an orthogonal frequency-division multiple access (OFDMA) framework exploiting the frequency-selective behavior of PASS \cite{OikonomouOFDMA2025}.

\subsection{Motivation and Contributions}

Despite the above progress, the user-specific beamforming gain of existing multi-user PASS designs essentially relies on PA-position optimization, which faces a fundamental timescale mismatch in time-division multiple access (TDMA) systems: mechanically repositioning PAs for each scheduled user requires a duration far exceeding a TDMA slot, whereas a fixed deployment cannot provide user-specific phase-coherent combining. Constrained pinching-antenna arrays (C-PAAs) improve deployment practicality by grouping multiple PAs into a compact movable module \cite{HouPinchingAntennaArray2026}, and it was further shown in \cite{JinCPAASumRate2026} that constrained PA movement can closely approach the ideal array-performance bound; nevertheless, slot-level user switching therein still entails fine PA-position adjustment inside the array. A natural question therefore arises: \emph{can PASS provide user-specific coherent combining while keeping all PA positions strictly fixed during user switching?}

This paper answers the above question affirmatively by introducing the carrier frequency as a fast slot-level control variable. The idea is rooted in the inherently frequency-dependent nature of waveguide-based radiation, as exemplified by the frequency-scanning behavior of leaky-wave antennas \cite{JacksonLWA2012, SarrazinLWADoA2026} and the frequency selectivity of PASS \cite{OikonomouOFDMA2025}. Such frequency dependence is either compensated as beam squint in wideband phased arrays \cite{WangBeamSquintTSP2019} or exploited via per-element frequency increments in frequency diverse arrays \cite{WangFDA2015}. In contrast, the proposed FreqTune-PASS applies a single user-specific carrier-frequency offset at the waveguide input, which exploits the distinct total delays of the PA-radiated paths to reshape their coherent superposition at the scheduled user even though all PA positions remain fixed. The main contributions of this paper are summarized as follows:

\begin{itemize}
    \item We propose FreqTune-PASS, a fixed-PA multi-user framework with slot-level frequency tuning, and develop its delay-domain system model that jointly captures the in-waveguide and free-space propagation delays of each PA-user path. Based on this model, the fast-timescale frequency-offset optimization and the slow-timescale PA-deployment problems are formulated.
    \item We characterize the coherent-combining mechanism of frequency tuning by expanding the received power into pairwise coherent terms, which reveals that the delay differences translate the frequency offset into distinct phase rotations, and demonstrate that phase alignment dominates the gain for uniformly distributed users.
    \item We derive a closed-form globally optimal offset for the two-PA case, develop a low-complexity one-dimensional search for the multi-PA case, and obtain an approximate closed-form offset under equal-spacing deployment via a Dirichlet-kernel interpretation.
    \item We study the long-term PA deployment under a finite tuning range and show that full-aperture equal spacing provides a robust delay structure for uniformly distributed users. Numerical results verify substantial sum-rate and worst-user gains over fixed-frequency transmission and show that the achieved rate approaches the coherent-combining upper bound.
\end{itemize}

\subsection{Organization and Notation}

The remainder of this paper is organized as follows. Section~II presents the system model and formulates the slot-level frequency-offset and long-term PA-deployment problems. Section~III develops the frequency-tuned beamforming design. Section~IV studies the corresponding long-term PA deployment design. Section~V provides numerical results. Finally, Section~VI concludes this paper.

Notation: Boldface letters denote vectors, where uppercase boldface letters are used for position vectors. $\mathbb{E}[\cdot]$ denotes statistical expectation, $\mathcal{CN}(\mu,\sigma^2)$ denotes a circularly symmetric complex Gaussian distribution with mean $\mu$ and variance $\sigma^2$, $\mathbb{Z}$ denotes the set of integers, and $\|\cdot\|$ denotes the Euclidean norm. The operator $\operatorname{dist}(x,\mathbb{Z})$ denotes the distance from a real number $x$ to its nearest integer, i.e.,  $\operatorname{dist}(x,\mathbb{Z})=\min_{k\in\mathbb{Z}}|x-k|$.

\section{System Model}
\label{sec:system_model}

\subsection{Antenna and Channel Model}
\label{subsec:antenna_channel_model}

\begin{figure}[!t]
\centering
\includegraphics[width=0.9\linewidth]{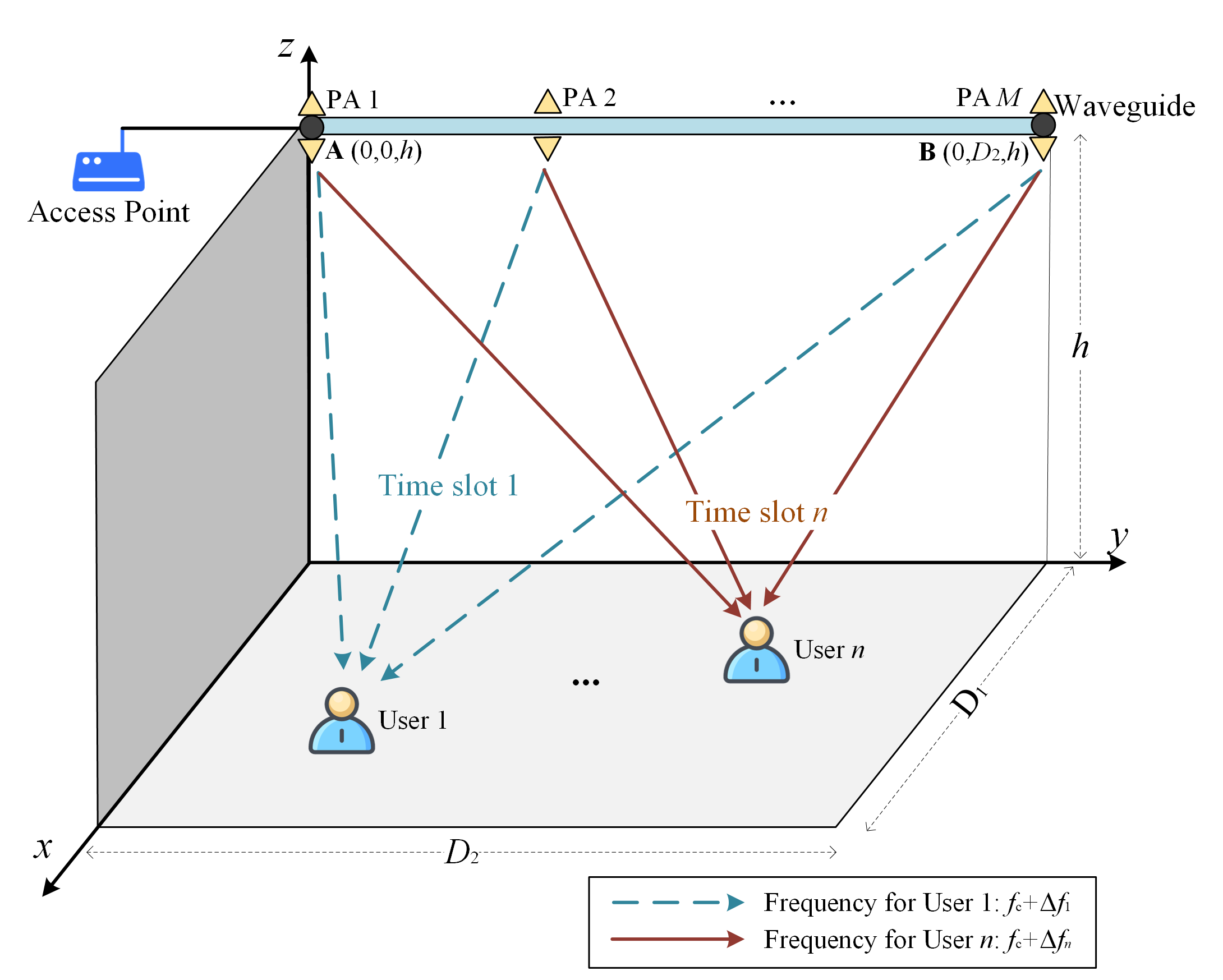}
\caption{Illustration of the considered single-waveguide PASS with frequency-tuned multi-PA coherent combining.}
\label{fig:Illustration_FTBO}
\end{figure}

As shown in Fig.~\ref{fig:Illustration_FTBO}, we consider a downlink pinching-antenna system deployed in an indoor room with dimensions $D_1 \times D_2 \times h$, where $D_1$, $D_2$, and $h$ denote the room length, width, and ceiling height, respectively. A dielectric waveguide is installed along one edge of the ceiling and is parallel to the $y$-axis. Without loss of generality, the waveguide extends from $\mathbf{A}=[0,0,h]$ to $\mathbf{B}=[0,D_2,h]$.

A total of $M$ PAs are deployed on the waveguide. The position of the $m$-th PA is denoted by
\begin{equation}
    \mathbf{Q}_m=[0,\Tilde{y}_m,h],
    \quad
    m=1,\ldots, M,
\end{equation}
where $0\leq \Tilde{y}_1\leq \cdots \leq \Tilde{y}_M\leq D_2$. $N$ single-antenna users distributed on the floor plane are served by the PASS. The location of user $n$ is given by
\begin{equation}
    \mathbf{U}_n=[x_n,y_n,0],
    \quad
    n=1,\ldots, N,
\end{equation}
where $x_n \in [0,D_1]$ and $y_n \in [0,D_2]$.

The PASS operates at a reference carrier frequency $f_c$. Due to the limited mechanical reconfiguration speed of PA positions, repositioning the PAs for user-specific beamforming may require a time duration far exceeding a TDMA slot. Therefore, the PAs remain static with fixed positions during slot and user switching, while a frequency offset $\Delta f_n$ is applied when serving user $n$ to provide fast slot-level signal alignment. The effective carrier frequency is denoted as
\begin{equation}
    f_n=f_c+\Delta f_n,
    \quad
    \Delta f_n\in[-\Delta f_{\max},\Delta f_{\max}].
\end{equation}
The corresponding free-space and guided wavelengths are given by
\begin{equation}
\label{eq:free_space_wavelength}
    \lambda_n=\frac{c}{f_n},
\end{equation}
and
\begin{equation}
\label{eq:guided_wavelength}
    \lambda_{g,n}=\frac{\lambda_n}{n_{\mathrm{eff}}},
\end{equation}
where $c$ is the speed of light, and $n_{\mathrm{eff}}$ denotes the effective refractive index of the waveguide.

The access point (AP) feeds the signal into the waveguide from $\mathbf{A}$. Before being radiated by the $m$-th PA, the signal propagates along the waveguide over a distance $\Tilde{y}_m$. Hence, the waveguide propagation response under frequency $f_n$ is given by
\begin{equation}
h_{m,n}^{\mathrm{wg}}(\Delta f_n)
    =
    \exp\left(
    -j\frac{2\pi \Tilde{y}_m}{\lambda_{g,n}}
    \right).
\end{equation}
After radiation, the signal propagates through free space to the intended user. Under the spherical-wave line-of-sight (LoS) model, the free-space channel coefficient between the $m$-th PA and user $n$ is 
\begin{equation}
    h_{m,n}^{\mathrm{fs}}(\Delta f_n)
    =
    \frac{\sqrt{\gamma_n(\Delta f_n)}}{d_{m,n}}
    \exp\left(
    -j\frac{2\pi d_{m,n}}{\lambda_n}
    \right),
\end{equation}
where $d_{m,n}$ is the distance between the $m$-th PA and user $n$, which can be computed as
\begin{equation}
    d_{m,n}=\|\mathbf{U}_n-\mathbf{Q}_m\|=\sqrt{x_n^2+(y_n-\Tilde{y}_m)^2+h^2}.
\end{equation}
$\gamma_n(\Delta f_n)$ is the frequency-dependent free-space path-loss coefficient at the reference distance, which can be expressed as
\begin{equation}
    \gamma_n(\Delta f_n)
    =
    \left(
    \frac{c}{4\pi(f_c+\Delta f_n)}
    \right)^2.
\end{equation}
Define the total propagation delay from the waveguide input to user $n$ through the $m$-th PA as
\begin{equation}
\label{eq:tau_mn}
    \tau_{m,n}
    =
    \frac{d_{m,n}}{c}
    +
    \frac{n_{\mathrm{eff}}\Tilde{y}_m}{c}.
\end{equation}
Combining the waveguide propagation phase and the free-space LoS channel, the equivalent channel from the waveguide input to user $n$ through the $m$-th PA is
\begin{equation}
    g_{m,n}(\Delta f_n)
    =
    h_{m,n}^{\mathrm{fs}}(\Delta f_n)
    h_{m,n}^{\mathrm{wg}}(\Delta f_n).
\end{equation}
Substituting the above expressions gives
\begin{equation}
\label{eq:equivalent_channel}
    g_{m,n}(\Delta f_n)
    =
    \frac{\sqrt{\gamma_n(\Delta f_n)}}{d_{m,n}}
    e^{-j2\pi(f_c+\Delta f_n)\tau_{m,n}}.
\end{equation}
It can be observed from \eqref{eq:equivalent_channel} that the frequency offset changes both the waveguide and the free-space propagation phase. Hence, even with fixed PA positions, frequency tuning provides an additional degree of freedom to reshape the coherent superposition of the multi-PA signals at the intended user.

\subsection{Signal Model}
\label{subsec:signal_model}

TDMA is adopted as a representative orthogonal multiple access protocol. In the slot assigned to user $n$, the AP transmits the data symbol $s_n$ intended for user $n$, where $\mathbb{E}\{|s_n|^2\}=1$. 
The total transmit power is denoted by $P$, and equal power allocation among the $M$ PAs is assumed, with the in-waveguide propagation loss neglected as in \cite{DingFlexiblePASS2025,OuyangArrayGainPASS2025}.
Accordingly, the received signal at user $n$ is
\begin{equation}
\label{eq:received_signal}
    r_n
    =
    \sum_{m=1}^{M}
    \sqrt{\frac{P}{M}}
    g_{m,n}(\Delta f_n)s_n
    +
    z_n,
\end{equation}
where $z_n \sim \mathcal{CN}(0,\sigma^2)$ denotes the additive white Gaussian noise (AWGN) at user $n$.

Based on \eqref{eq:received_signal}, the received signal-to-noise ratio (SNR) of user $n$ is expressed as
\begin{equation}
\label{eq:snr}
    \Gamma_n(\Delta f_n)
    =
    \frac{P}{M\sigma^2}
    \left|
    \sum_{m=1}^{M}
    g_{m,n}(\Delta f_n)
    \right|^2 .
\end{equation}
The achievable rate of user $n$ under equal TDMA time sharing is given by
\begin{equation}
\label{eq:user_rate}
    R_n(\Delta f_n)
    =
    \frac{1}{N}
    \log_2
    \left(
    1+
    \Gamma_n(\Delta f_n)
    \right),
\end{equation}
where the pre-log factor $1/N$ accounts for the equal time sharing among the $N$ users. The corresponding system sum rate can be computed as
\begin{equation}
\label{eq:sum_rate}
    R_{\mathrm{sum}}
    =
    \sum_{n=1}^{N}
    R_n(\Delta f_n).
\end{equation}

\subsection{Problem Formulation}
\label{subsec:problem_formulation}

As discussed in Section~\ref{subsec:antenna_channel_model}, the PA positions and the frequency offsets operate on two different timescales: the former serve as a long-term deployment variable designed according to the user distribution, whereas the latter is updated at the slot level for the currently scheduled user.

For a given PA deployment $\mathbf y=[\Tilde{y}_1,\ldots,\Tilde{y}_M]^T$, the slot-level frequency-offset optimization for user $n$ is formulated as
\begin{subequations}
\label{prob:slot_frequency_design}
\begin{align}
    \max_{\Delta f_n} \quad
    & R_n(\Delta f_n) \\
    \mathrm{s.t.}\quad
    & -\Delta f_{\max}
    \leq
    \Delta f_n
    \leq
    \Delta f_{\max}.
\end{align}
\end{subequations}
Since TDMA serves one user per slot, \eqref{prob:slot_frequency_design} can be solved independently for each user under a fixed PA deployment.

The long-term PA deployment should be suitable for randomly distributed users rather than for a particular instantaneous user realization. For a generic scheduled user located at $\mathbf {U}_n$, the corresponding statistical deployment problem can be written as
\begin{subequations}
\label{prob:statistical_deployment}
\begin{align}
    \max_{\mathbf y} \quad
    &
    \mathbb{E}_{\mathbf U_n}
    \left[
    \max_{|\Delta f_n|\leq \Delta f_{\max}}
    R_n(\mathbf U_n,\mathbf y,\Delta f_n)
    \right] \\
    \mathrm{s.t.}\quad
    &
    0\leq \Tilde{y}_1\leq \cdots \leq \Tilde{y}_M\leq D_2, \\
    &
    \Tilde{y}_{m+1}-\Tilde{y}_m\geq \Delta_{\min},
    \quad m=1,\ldots,M-1 ,
\end{align}
\end{subequations}
where the expectation is taken over the random scheduled-user location, and $\Delta_{\min}$ denotes the minimum PA spacing required by mutual coupling and practical PA-placement considerations \cite{OuyangArrayGainPASS2025}. In this paper, $\Delta_{\min}$ is chosen no smaller than $\lambda_c/2$, where $\lambda_c=c/f_c$ is the wavelength at the reference carrier frequency.

Problem~\eqref{prob:statistical_deployment} is challenging because the expectation contains the slot-level optimization in \eqref{prob:slot_frequency_design} as an inner problem. Moreover, the PA positions affect both the distance-dependent amplitudes and the pairwise propagation phases, leading to a highly nonconvex statistical objective with respect to $\mathbf y$. Therefore, we first solve \eqref{prob:slot_frequency_design} for an arbitrary fixed PA deployment, which reveals how the finite frequency-offset range controls the pairwise phase alignment. Based on this result, the long-term deployment problem in \eqref{prob:statistical_deployment} is then revisited and simplified into a phase-tuning-capability design for uniformly distributed users.

\section{Frequency-Tuned Beamforming}
\label{sec:frequency_tuned_phase_alignment}

This section studies the slot-level frequency-offset optimization for a given fixed PA deployment. The PA positions are treated as slow-timescale variables, while the frequency offset is updated for the scheduled user in each TDMA slot. We first characterize the pairwise coherent-combining structure and then solve the frequency-offset optimization problem for the two-PA and multi-PA cases.

\subsection{Physical Interpretation of Frequency-Tuned Beamforming}
\label{subsec:physical_interpretation_ft_beamforming}

In conventional antenna arrays, beamforming is usually realized by adjusting the complex weight of each antenna element. The considered PASS follows a different principle. The PA positions remain fixed during TDMA slot switching, and no fast per-PA phase shifter is assumed. Instead, the serving frequency is slightly tuned for the scheduled user. Although the same frequency offset is applied to all PA-radiated paths, it changes their accumulated phases by different amounts because the total propagation delay is different for each PA path.

For user $n$ located at an arbitrary position in the service region, the frequency-dependent coherent-combining gain is characterized by the weighted factor
\begin{equation}
\label{eq:exact_weighted_factor}
    A_{w,n}(\Delta f_n)
    =
    \left|
    \sum_{m=1}^{M}
    a_{m,n}
    e^{-j2\pi(f_c+\Delta f_n)\tau_{m,n}}
    \right|^2,
\end{equation}
where $a_{m,n}=1/d_{m,n}$ is the distance-dependent amplitude coefficient, and $\tau_{m,n}$ is the total delay from the waveguide input to user $n$ through PA $m$. When $\Delta f_n$ changes, the phase of the path through PA $m$ is rotated by $2\pi\Delta f_n\tau_{m,n}$. Since $\tau_{m,n}$ varies with the PA index and the user location, the same frequency offset produces different phase rotations across the PA-radiated paths. 
As a result, frequency tuning changes their constructive or destructive superposition at user $n$ and reshapes the spatial power pattern in the indoor service region. Therefore, frequency-tuned beamforming in this paper refers to spatial power shaping through frequency control, rather than conventional beamforming with independent per-antenna weights.

\begin{figure}[!t]
\centering
\includegraphics[width=1\linewidth]{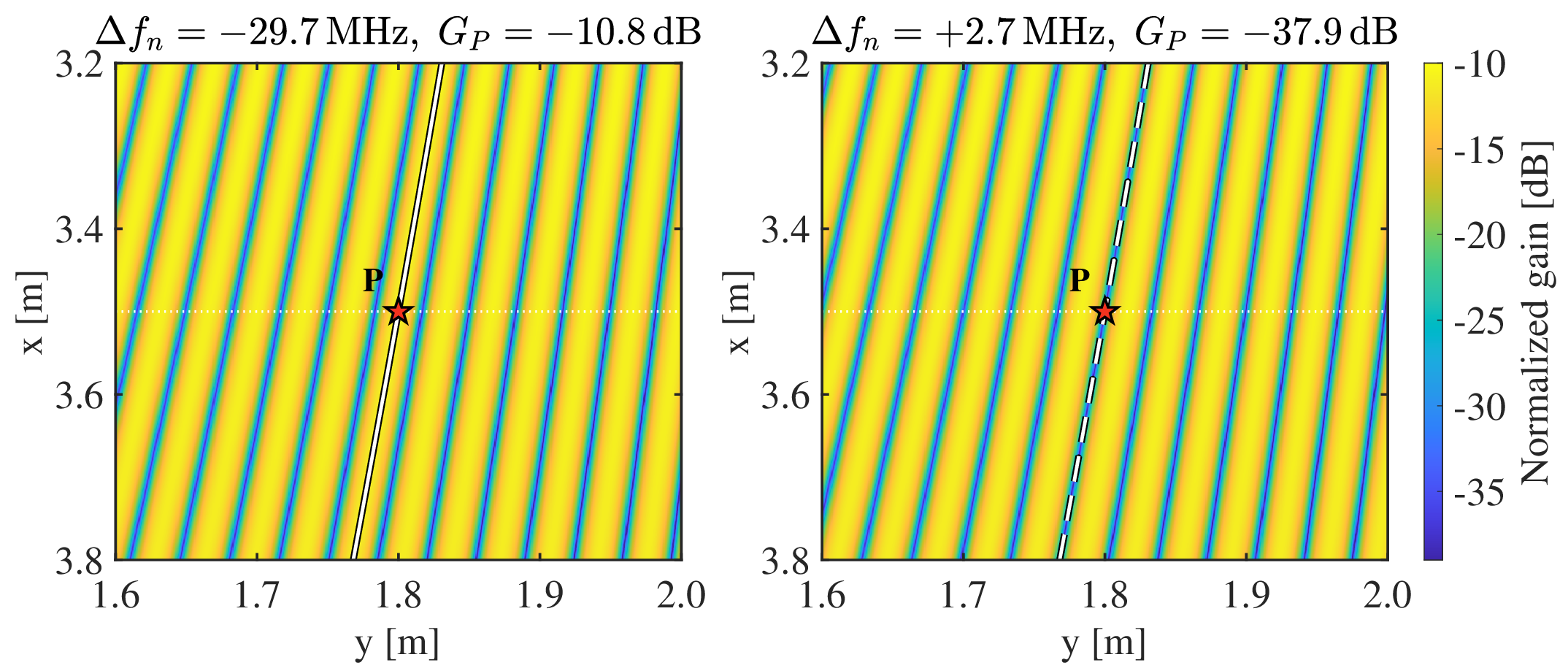}
\caption{Normalized coherent-combining gain $A_{w,n}(\Delta f_n)/M$ under two frequency offsets.}
\label{fig:freq_tuned_peak_valley_map}
\end{figure}

To further illustrate the physical meaning of frequency-tuned beamforming, 
Fig.~\ref{fig:freq_tuned_peak_valley_map} considers a representative two-PA setup in a $5~\mathrm{m}\times5~\mathrm{m}\times3~\mathrm{m}$ room with $f_c=15$ GHz and $n_{\mathrm{eff}}=1.4$, where the two PAs are fixed at $\mathbf Q_1=[0,1,3]$ and $\mathbf Q_2=[0,4,3]$ on the ceiling-edge waveguide, and the observation point is $\mathbf P=[3.5,1.8,0]$. The plotted quantity is the normalized coherent-combining gain $A_{w,n}(\Delta f_n)/M$, which excludes the nearly constant free-space factor $\gamma_n(\Delta f_n)$, and only the frequency offset is varied.
As shown in Fig.~\ref{fig:freq_tuned_peak_valley_map}, the offset $\Delta f_n=-29.7$ MHz places $\mathbf P$ near a constructive-combining peak, whereas $\Delta f_n=+2.7$ MHz places the same point near a destructive-combining valley. The corresponding normalized gains are $-10.8$ dB and $-37.9$ dB, respectively, which yields a $27.1$ dB gain difference.
This example shows that frequency tuning can shift the peak-valley pattern and substantially change the received power at a fixed spatial location.
\begin{figure}[!t]
\centering
\includegraphics[width=0.9\linewidth]{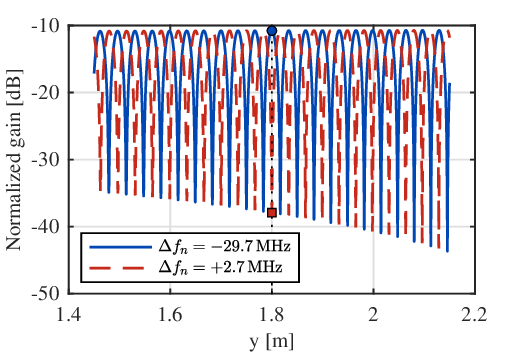}
\caption{Normalized gain profiles along $x=3.5$ m.}
\label{fig:freq_tuned_peak_valley_profile}
\end{figure}

Fig.~\ref{fig:freq_tuned_peak_valley_profile} further plots the normalized gain along the line $x=3.5$ m under the same two frequency offsets. The two curves exhibit a clear peak-valley conversion: the frequency offset that creates a gain peak at a certain location can be changed to make the same location fall into a gain valley, and vice versa. Therefore, frequency tuning does not merely scale the received power uniformly; instead, it shifts the spatial interference pattern generated by the fixed PAs and enables peak-valley switching over the service region. This is the key mechanism behind frequency-tuned beamforming in PASS. For TDMA-based multi-user service, the AP can select a user-specific frequency offset in each slot so that the scheduled user is moved closer to a constructive-combining peak without physically relocating the PAs.

\subsection{Pairwise Coherent-Combining Structure}
\label{subsec:pairwise_coherent_combining}

The frequency-tuned beamforming effect in \eqref{eq:exact_weighted_factor} can be further understood through its pairwise coherent-combining structure. Expanding $A_{w,n}(\Delta f_n)$ yields
\begin{equation}
\label{eq:weighted_pairwise_expansion}
    A_{w,n}
    =
    \sum_{m=1}^{M}a_{m,n}^{2}
    +
    2\sum_{1\leq m<\ell\leq M}
    a_{m,n}a_{\ell,n}
    \cos \phi_{\ell,m,n},
\end{equation}
where $\phi_{\ell,m,n}$ is the pairwise phase difference between the signals radiated by the $\ell$-th PA and the $m$-th PA at user $n$, which can be computed as
\begin{equation}
\label{eq:pairwise_phase}
    \phi_{\ell,m,n}
    =
    2\pi(f_c+\Delta f_n)
    \left(
    \tau_{\ell,n}-\tau_{m,n}
    \right).
\end{equation}
Equation~\eqref{eq:weighted_pairwise_expansion} shows that the received power contains $M(M-1)/2$ pairwise coherent terms. When these terms are phase-aligned, the cosine terms approach one and the PA-assisted paths are coherently combined. When the phases are mismatched, the cross terms may vanish or become negative, leading to a substantial received-power loss.

\noindent\textbf{Lemma 1:}
Consider two PA-radiated paths from PA $m$ and PA $\ell$ to user $n$. When the serving frequency is changed from the reference carrier $f_c$ to $f_c+\Delta f_n$, the relative phase between these two paths is governed by their delay difference $\tau_{\ell,n}-\tau_{m,n}$. Under the frequency-offset constraint $|\Delta f_n|\leq \Delta f_{\max}$, the maximum tunable relative-phase span is
\begin{equation}
\label{eq:pair_phase_span}
    \Omega_{\ell,m,n}
    =
    4\pi\Delta f_{\max}
    \left|
    \tau_{\ell,n}-\tau_{m,n}
    \right|.
\end{equation}
Thus, the phase-tuning capability of a PA pair is determined by the available frequency-offset range and the relative delay between the two PA paths.

\noindent\textit{Proof:}
From \eqref{eq:pairwise_phase}, the phase variation induced by $\Delta f_n$ relative to the reference carrier frequency is
\begin{equation}
\label{eq:tunable_pairwise_phase}
    \Delta\phi_{\ell,m,n}^{\mathrm{tune}}
    =
    \phi_{\ell,m,n}(\Delta f_n)-\phi_{\ell,m,n}(0)
    =
    2\pi\Delta f_n
    \left(
    \tau_{\ell,n}-\tau_{m,n}
    \right).
\end{equation}
Since $\Delta f_n\in[-\Delta f_{\max},\Delta f_{\max}]$, the maximum excursion of $\Delta\phi_{\ell,m,n}^{\mathrm{tune}}$ over the feasible interval is
\begin{equation}
    \max_{\Delta f_n}
    \Delta\phi_{\ell,m,n}^{\mathrm{tune}}
    -
    \min_{\Delta f_n}
    \Delta\phi_{\ell,m,n}^{\mathrm{tune}}
    =
    4\pi\Delta f_{\max}
    \left|
    \tau_{\ell,n}-\tau_{m,n}
    \right|,
\end{equation}
which completes the proof.

\noindent\textbf{Remark 1:}
For given PA positions and a given user location, the relative delay $\tau_{\ell,n}-\tau_{m,n}$ is fixed. Lemma~1 shows that the tunable phase span is determined by the absolute frequency-offset range $\Delta f_{\max}$ and the relative delay, which is independent of the reference carrier frequency $f_c$. Therefore, the same $\Delta f_{\max}$ provides the same phase-tuning capability at different reference carrier frequencies, which indicates that frequency-offset-based beamforming remains effective in high-frequency communication systems: even when the relative offset $\Delta f_{\max}/f_c$ becomes small, the absolute offset $\Delta f_{\max}$ can provide the same controllable phase span for the same propagation-delay difference.

\subsection{Slot-Level Frequency-Offset Optimization}
\label{subsec:slot_level_frequency_optimization}

For a fixed PA deployment, the slot-level design problem has been formulated in \eqref{prob:slot_frequency_design}. Because the user rate is monotonically increasing with the received SNR and the TDMA pre-log factor is independent of $\Delta f_n$, this slot-level design can be reduced to SNR maximization.
Since $|\Delta f_n|\ll f_c$, the variation of $\gamma_n(\Delta f_n)$ over the feasible interval is negligible compared with the phase variation. Therefore, the offset design is guided by maximizing the weighted coherent-combining factor:
\begin{subequations}
\label{prob_weighted_frequency}
\begin{align}
    \max_{\Delta f_n} \quad
    & A_{w,n}(\Delta f_n) \\
    \mathrm{s.t.}\quad
    & -\Delta f_{\max}
    \leq
    \Delta f_n
    \leq
    \Delta f_{\max}.
\end{align}
\end{subequations}
Hereinafter, the optimality of the frequency offset is stated with respect to \eqref{prob_weighted_frequency}.
The structure of $A_{w,n}(\Delta f_n)$ in \eqref{eq:weighted_pairwise_expansion} explains why the two-PA and multi-PA cases should be treated separately. When $M=2$, $A_{w,n}(\Delta f_n)$ contains only one pairwise phase term. Since the amplitude weights $a_{1,n}$ and $a_{2,n}$ are independent of $\Delta f_n$ except for the negligible common free-space loss factor, maximizing $A_{w,n}(\Delta f_n)$ reduces to aligning a single cosine term. This special structure reduces the two-PA offset design to a single phase-cycle matching problem, which leads to a closed-form optimal frequency offset.
When $M>2$, however, $A_{w,n}(\Delta f_n)$ contains $M(M-1)/2$ pairwise cosine terms. These terms have different delay coefficients $\tau_{\ell,n}-\tau_{m,n}$, while only one scalar frequency offset $\Delta f_n$ is available. Therefore, increasing one cosine term may decrease another, and the perfect-alignment conditions become a set of generally inconsistent integer equations.
\subsubsection{Two-PA Closed-Form Solution}

For $M=2$, define the absolute delay difference between the two PA paths as
\begin{equation}
\label{eq:two_pa_eta}
    \eta_n=
    |\tau_{2,n}-\tau_{1,n}|,
\end{equation}
which determines the phase sensitivity of the two-PA coherent-combining term to the frequency offset.
Then \eqref{eq:weighted_pairwise_expansion} becomes
\begin{equation}
\label{eq:two_pa_weighted_factor}
    A_{w,n}(\Delta f_n)
    =
    a_{1,n}^2+a_{2,n}^2
    +
    2a_{1,n}a_{2,n}
    \cos\left(
    2\pi(f_c+\Delta f_n)\eta_n
    \right).
\end{equation}

\noindent\textbf{Proposition 1:}
Consider the two-PA case for user $n$ with the absolute delay difference $\eta_n>0$. Define the feasible normalized phase-cycle interval as
\begin{equation}
\label{eq:two_pa_interval}
    \mathcal I_n
    =
    \left[
    (f_c-\Delta f_{\max})\eta_n,
    (f_c+\Delta f_{\max})\eta_n
    \right].
\end{equation} 
If $\mathbb Z\cap\mathcal I_n\neq\emptyset$, perfect phase alignment is feasible, and the globally optimal frequency offset is
\begin{equation}
\label{eq:two_pa_closed_form}
    \Delta f_n^\star
    =
    \frac{k_n^\star}{\eta_n}-f_c,
\end{equation}
where $k_n^\star$ can be any integer in $\mathbb Z\cap\mathcal I_n$. If $\mathbb Z\cap\mathcal I_n=\emptyset$, perfect phase alignment is infeasible, and the globally optimal frequency offset is
\begin{equation}
\label{eq:two_pa_endpoint}
    \Delta f_n^\star
    =
    \operatorname*{arg\,min}_{\Delta f_n\in\mathcal B}
    \operatorname{dist}
    \left(
    (f_c+\Delta f_n)\eta_n,\mathbb Z
    \right),
\end{equation}
where $\mathcal B = \{-\Delta f_{\max},\Delta f_{\max}\}$.

\noindent\textit{Proof:}
Since $a_{1,n}a_{2,n}>0$, maximizing \eqref{eq:two_pa_weighted_factor} is equivalent to maximizing
$\cos\left(2\pi(f_c+\Delta f_n)\eta_n\right)$.
The cosine term is maximized when $(f_c+\Delta f_n)\eta_n\in\mathbb Z$.
Because $\Delta f_n\in[-\Delta f_{\max},\Delta f_{\max}]$, the feasible range of $(f_c+\Delta f_n)\eta_n$ is $\mathcal I_n$. If $\mathbb Z\cap\mathcal I_n\neq\emptyset$, \eqref{eq:two_pa_closed_form} follows directly.

If $\mathbb Z\cap\mathcal I_n=\emptyset$, perfect alignment is infeasible. Since $\mathcal I_n$ is an interval containing no integer, the minimum distance to $\mathbb Z$ is attained at one of its endpoints, which corresponds to $\Delta f_n\in\mathcal B$. This proves \eqref{eq:two_pa_endpoint}.

\subsubsection{Multi-PA One-Dimensional Search}

For $M>2$, one scalar frequency offset must jointly tune multiple pairwise phases. Perfect phase alignment requires
\begin{equation}
\label{eq:perfect_alignment_condition}
    (f_c+\Delta f_n)
    \left(
    \tau_{m,n}-\tau_{1,n}
    \right)
    =
    k_{m,n},
    \,
    k_{m,n}\in\mathbb Z,
    \,
    m=2,\ldots,M.
\end{equation}
These integer conditions are generally inconsistent because the relative delays are user-dependent and cannot, in general, be represented as integer multiples of a common delay unit. Hence, a universal closed-form global optimum is generally unavailable for arbitrary PA deployments.

Nevertheless, \eqref{prob_weighted_frequency} is a one-dimensional optimization problem. Differentiating the pairwise form of $A_{w,n}(\Delta f_n)$ gives
\begin{multline}
\label{eq:weighted_derivative}
    A_{w,n}'(\Delta f_n)
    =
    -4\pi
    \sum_{1\leq m<\ell\leq M}
    a_{m,n}a_{\ell,n}
    (\tau_{\ell,n}-\tau_{m,n})\\
    {}\times
    \sin\left(
    2\pi(f_c+\Delta f_n)
    (\tau_{\ell,n}-\tau_{m,n})
    \right).
\end{multline}

\noindent\textbf{Proposition 2:}
For the multi-PA case with a fixed PA deployment and a given user $n$, the global optimizer of \eqref{prob_weighted_frequency} must belong to the candidate set
\begin{equation}
\label{eq:candidate_set}
    \mathcal C_n
    =
    \{-\Delta f_{\max},\Delta f_{\max}\}
    \cup
    \{\Delta f_n:A_{w,n}'(\Delta f_n)=0\}.
\end{equation}
Therefore, the optimal offset can be obtained by evaluating the exact objective over this candidate set:
\begin{equation}
\label{eq:multi_pa_solution}
    \Delta f_n^\star
    =
    \operatorname*{arg\,max}_{\Delta f_n\in\mathcal C_n}
    A_{w,n}(\Delta f_n).
\end{equation}

\noindent\textit{Proof:}
The feasible set $[-\Delta f_{\max},\Delta f_{\max}]$ is compact, and $A_{w,n}(\Delta f_n)$ is continuous and differentiable with respect to $\Delta f_n$. Hence, a global maximum exists. If it is attained at an interior point, Fermat's rule gives $A_{w,n}'(\Delta f_n)=0$; otherwise, it is attained at one of the two boundaries. This completes the proof. 

Although Proposition~2 characterizes the candidate set containing the global optimum, the stationary points in $\mathcal C_n$ are roots of $A_{w,n}'(\Delta f_n)=0$. From \eqref{eq:weighted_derivative}, this equation is a weighted sum of sinusoids with distinct delay coefficients and thus admits no closed-form enumeration.
Hence, we develop Algorithm~\ref{alg:multi_pa_frequency} to solve \eqref{prob_weighted_frequency} through a grid-assisted one-dimensional search over the feasible frequency-offset interval.
The basic idea of this algorithm is to first sample the feasible frequency-offset interval, then refine the grid intervals containing sign changes of $A_{w,n}'(\Delta f_n)$ to locate stationary points, and select the candidate maximizing the exact objective.
If $K$ stationary intervals are refined with $I$ iterations, the computational complexity is $\mathcal O(QM^2+KIM^2)$ per user, since evaluating $A_{w,n}(\Delta f_n)$ or $A_{w,n}'(\Delta f_n)$ involves all PA pairs. The optimization is performed independently across TDMA slots.

\begin{algorithm}[!t]
\caption{One-Dimensional Frequency-Offset Search}
\label{alg:multi_pa_frequency}
\begin{algorithmic}[1]
\REQUIRE $f_c$, $\Delta f_{\max}$, $\{\tau_{m,n}\}_{m=1}^{M}$, $\{a_{m,n}\}_{m=1}^{M}$, grid size $Q$.
\ENSURE $\Delta f_n^\star$.
\STATE Generate a uniform grid $\mathcal G=\{\nu_q\}_{q=1}^{Q}$ over $[-\Delta f_{\max},\Delta f_{\max}]$.
\STATE Initialize $\widehat{\mathcal C}_n=\mathcal G$.
\FOR{$q=1,\ldots,Q-1$}
    \IF{$A_{w,n}'(\nu_q)A_{w,n}'(\nu_{q+1})\leq0$}
        \STATE Refine a stationary point in $[\nu_q,\nu_{q+1}]$ by bisection or Brent's method.
        \STATE Add the refined point to $\widehat{\mathcal C}_n$.
    \ENDIF
\ENDFOR
\STATE Obtain $\Delta f_n^\star=\operatorname*{arg\,max}_{\Delta f_n\in\widehat{\mathcal C}_n}A_{w,n}(\Delta f_n)$.
\RETURN $\Delta f_n^\star$.
\end{algorithmic}
\end{algorithm}

\section{PA Deployment Design for Frequency-Tuned Beamforming}
\label{sec:pa_deployment_frequency_tuned}

Since frequency tuning controls the relative phases through the PA-user delay differences, the PA positions directly determine the beamforming capability achievable under a finite frequency-offset range. 
Therefore, this section investigates long-term PA deployment designs that provide robust delay structures for randomly distributed users, thereby supporting reliable indoor service.

\subsection{Deployment Metrics Under Finite Frequency Tuning}
\label{subsec:deployment_metrics_finite_tuning}

Let $L$ denote the waveguide length. For simplicity, this paper assumes that the waveguide spans the entire room width, i.e., $L=D_2$.
For a generic user location $\mathbf U=[x,y,0]$, define the continuous delay function along the waveguide as
\begin{equation}
\label{eq:tau_continuous_y}
    \tau(\Tilde{y};\mathbf U)
    =
    \frac{\sqrt{x^2+(y-\Tilde{y})^2+h^2}}{c}
    +
    \frac{n_{\mathrm{eff}}\Tilde{y}}{c},
    \quad
    \Tilde{y}\in[0,L].
\end{equation}
For adjacent PAs, define the spacing
\begin{equation}
\label{eq:pa_spacing}
    g_m
    =
    \Tilde{y}_{m+1}-\Tilde{y}_m,
    \quad
    m=1,\ldots,M-1,
\end{equation}
and the adjacent delay increment
\begin{equation}
\label{eq:adjacent_delay_increment}
    \delta_m(\mathbf U)
    =
    \tau(\Tilde{y}_{m+1};\mathbf U)-\tau(\Tilde{y}_m;\mathbf U).
\end{equation}
The pairwise delay difference between any two PAs can be expressed as
\begin{equation}
\label{eq:pairwise_delay_sum}
    \tau(\Tilde{y}_\ell;\mathbf U)-\tau(\Tilde{y}_m;\mathbf U)
    =
    \sum_{i=m}^{\ell-1}
    \delta_i(\mathbf U),
    \quad
    \ell>m.
\end{equation}
Therefore, the multi-PA phase-alignment problem is governed by the adjacent delay increments. A good deployment should satisfy two requirements. First, each adjacent PA pair should have a sufficiently large tunable phase span under the finite frequency-offset range. Second, the adjacent delay increments should be as regular as possible, so that one scalar frequency offset can tune multiple pairwise phases coherently.

To quantify these requirements, we introduce the geometric parameter
\begin{equation}
\label{eq:kappa_D_definition}
    \kappa_D
    =
    \frac{D_2}{\sqrt{D_2^2+h^2}} < 1,
\end{equation}
which is the maximum direction cosine between a PA--user LoS path and the waveguide axis.

\noindent\textbf{Lemma 2:}
For any user location $\mathbf U$ and any adjacent PA pair $(m,m+1)$, if $n_{\mathrm{eff}}>\kappa_D$, the delay function $\tau(\Tilde{y};\mathbf U)$ is strictly increasing with $\Tilde{y}$, and the adjacent delay increment satisfies
\begin{equation}
\label{eq:adjacent_delay_bound}
    \frac{g_m}{c}\left(n_{\mathrm{eff}}-\kappa_D\right)
    \leq
    \delta_m(\mathbf U)
    \leq
    \frac{g_m}{c}\left(n_{\mathrm{eff}}+\kappa_D\right) .
\end{equation}
Consequently, the tunable phase span of the adjacent PA pair is lower bounded by
\begin{equation}
\label{eq:adjacent_phase_lower_bound}
    \Omega_{m+1,m}(\mathbf U)
    \geq
    \frac{4\pi\Delta f_{\max}(n_{\mathrm{eff}}-\kappa_D)}{c}
    g_m,
\end{equation}
where $\Omega_{m+1,m}(\mathbf U)$ denotes the tunable phase span in \eqref{eq:pair_phase_span} evaluated at the generic location $\mathbf U$.

\noindent\textit{Proof:}
The derivative of \eqref{eq:tau_continuous_y} is
\begin{equation}
\label{eq:tau_derivative}
    \frac{\partial \tau(\Tilde{y};\mathbf U)}{\partial \Tilde{y}}
    =
    \frac{\Tilde{y}-y}{c\sqrt{x^2+(y-\Tilde{y})^2+h^2}}
    +
    \frac{n_{\mathrm{eff}}}{c}.
\end{equation}
Since
\begin{equation}
\label{eq:normalized_slope_bound}
    -\kappa_D
    \leq
    \frac{\Tilde{y}-y}{\sqrt{x^2+(y-\Tilde{y})^2+h^2}}
    \leq
    \kappa_D,
\end{equation}
we have
\begin{equation}
\label{eq:tau_derivative_bound}
    \frac{n_{\mathrm{eff}}-\kappa_D}{c}
    \leq
    \frac{\partial \tau(\Tilde{y};\mathbf U)}{\partial \Tilde{y}}
    \leq
    \frac{n_{\mathrm{eff}}+\kappa_D}{c}.
\end{equation}
Since $n_{\mathrm{eff}}>\kappa_D$, the lower bound in \eqref{eq:tau_derivative_bound} is positive, and thus $\tau(\Tilde{y};\mathbf U)$ is strictly increasing with $\Tilde{y}$. 
Applying the mean value theorem to $\tau(\Tilde{y};\mathbf U)$ over $[\Tilde{y}_m,\Tilde{y}_{m+1}]$, there exists $\xi_m\in(\Tilde{y}_m,\Tilde{y}_{m+1})$ such that
\begin{equation}
\begin{split}
    \delta_m(\mathbf U)
    &=
    \tau(\Tilde{y}_{m+1};\mathbf U)-\tau(\Tilde{y}_m;\mathbf U)\\
    &=
    \left.
    \frac{\partial \tau(\Tilde{y};\mathbf U)}{\partial \Tilde{y}}
    \right|_{\Tilde{y}=\xi_m}
    (\Tilde{y}_{m+1}-\Tilde{y}_m).
\end{split}   
\end{equation}
Since $\Tilde{y}_{m+1}-\Tilde{y}_m=g_m$, substituting the derivative bound in \eqref{eq:tau_derivative_bound} gives \eqref{eq:adjacent_delay_bound}. Combining \eqref{eq:adjacent_delay_bound} with Lemma~1 gives \eqref{eq:adjacent_phase_lower_bound}. This completes the proof.

Note that $\kappa_D<1$ always holds, while $n_{\mathrm{eff}}>1$ for practical dielectric waveguides; hence, the condition $n_{\mathrm{eff}}>\kappa_D$ is automatically satisfied.
Lemma~2 shows that short adjacent PA spacings lead to weakly tunable adjacent phase differences. Therefore, in the multi-PA case, avoiding excessively small adjacent spacings is essential for maintaining robust joint phase alignment under a finite frequency-offset range.

\subsection{Full-Aperture Equal-Spacing Deployment}
\label{subsec:full_aperture_equal_spacing}

We next derive the PA deployment that is compatible with finite-offset frequency tuning. The derivation is based on two criteria: maximizing the guaranteed adjacent phase-tuning span and making the delay sequence close to an arithmetic progression.

\noindent\textbf{Proposition 3:}
Assume that $M$ PAs are deployed on a waveguide of length $L$ with spacing constraint $g_m\geq\Delta_{\min}$. If
\begin{equation}
\label{eq:uniform_feasibility}
    \frac{L}{M-1}
    \geq
    \Delta_{\min},
\end{equation}
then the deployment
\begin{equation}
\label{eq:uniform_pa_deployment}
    \Tilde{y}_m^{\mathrm{eq}}
    =
    \frac{(m-1)L}{M-1},
    \quad
    m=1,\ldots,M,
\end{equation}
maximizes the minimum adjacent spacing $\min_m g_m$ among all feasible deployments. Consequently, it maximizes the guaranteed lower bound of the worst adjacent phase-tuning span in \eqref{eq:adjacent_phase_lower_bound}.

\noindent\textit{Proof:}
For any feasible deployment,
\begin{equation}
\label{eq:min_spacing_bound}
    \min_m g_m
    \leq
    \frac{1}{M-1}
    \sum_{m=1}^{M-1}g_m
    =
    \frac{\Tilde{y}_M-\Tilde{y}_1}{M-1}
    \leq
    \frac{L}{M-1}.
\end{equation}
The upper bound is achieved if and only if $\Tilde{y}_1=0$, $\Tilde{y}_M=L$, and all adjacent spacings are equal to $L/(M-1)$. Under \eqref{eq:uniform_feasibility}, this deployment satisfies the minimum-spacing constraint. Therefore, \eqref{eq:uniform_pa_deployment} maximizes $\min_m g_m$ and hence maximizes the guaranteed lower bound in \eqref{eq:adjacent_phase_lower_bound}. This completes the proof.

The equal-spacing deployment is feasible under the minimum-spacing constraint if \eqref{eq:uniform_feasibility} holds. Equivalently, for a given waveguide length $L$ and minimum spacing $\Delta_{\min}$, the number of deployable PAs should satisfy
\begin{equation}
\label{eq:max_pa_feasibility}
    M
    \leq
    \left\lfloor
    \frac{L}{\Delta_{\min}}
    \right\rfloor
    +1.
\end{equation}
In high-frequency PASS, $\Delta_{\min}$ is typically on the order of half a wavelength, i.e., $\Delta_{\min}\geq \lambda_c/2$. Since $\lambda_c$ is at the centimeter or millimeter scale while the indoor waveguide length is usually at the meter scale, the minimum-spacing constraint is generally not the limiting factor for the considered deployment. Therefore, the equal-spacing deployment can usually satisfy the coupling-related spacing requirement while using the full waveguide aperture.

\noindent\textbf{Remark 2:}
For $M=2$, Proposition~3 reduces to the endpoint deployment, which is intuitive because only one relative phase needs to be tuned. For $M>2$, however, the deployment should also avoid highly uneven adjacent spacings. Although a nonuniform deployment may create a large separation for some PA pairs, excessively short adjacent spacings lead to weakly tunable phase differences and may become bottlenecks for joint phase alignment. Equal spacing avoids such weakly controllable adjacent PA pairs.

We now show that equal spacing is also favorable from the viewpoint of joint multi-PA phase tuning. Define the average adjacent delay increment as
\begin{equation}
\label{eq:average_adjacent_delay}
    \bar\delta(\mathbf U)
    =
    \frac{1}{M-1}
    \sum_{m=1}^{M-1}
    \delta_m(\mathbf U),
\end{equation}
and define the adjacent-delay dispersion as
\begin{equation}
\label{eq:adjacent_delay_dispersion}
    \mathcal V_{\delta}(\mathbf y,\mathbf U)
    =
    \sum_{m=1}^{M-1}
    \left(
    \delta_m(\mathbf U)-\bar\delta(\mathbf U)
    \right)^2 .
\end{equation}
A smaller $\mathcal V_{\delta}(\mathbf y,\mathbf U)$ means that the adjacent delay increments are more uniform. In this case, the relative delays are closer to an arithmetic progression, and the multiple phase-alignment conditions can be approximately reduced to one fundamental condition.

To characterize the delay regularity, we decompose the delay function $\tau(\Tilde{y};\mathbf U)$ as
\begin{equation}
\label{eq:delay_guided_free_space_decomposition}
    \tau(\Tilde{y};\mathbf U)
    =
    \frac{n_{\mathrm{eff}}\Tilde{y}}{c}
    +
    r(\Tilde{y};\mathbf U),
\end{equation}
where $r(\Tilde{y};\mathbf U) = \sqrt{x^2+(y-\Tilde{y})^2+h^2}/c$ is the nonlinear free-space propagation delay from the PA position to user location $\mathbf U$. Accordingly, the adjacent delay increment can be written exactly as
\begin{equation}
\label{eq:delta_guided_residual}
    \delta_m(\mathbf U)
    =
    \frac{n_{\mathrm{eff}}}{c}g_m
    +
    \zeta_m(\mathbf U),
\end{equation}
where $\zeta_m(\mathbf U) = r(\Tilde{y}_{m+1};\mathbf U)-r(\Tilde{y}_m;\mathbf U)$ is the free-space-delay perturbation over the $m$-th adjacent PA interval.

The perturbation term is bounded. By the mean value theorem, there exists $\xi_m\in(\Tilde{y}_m,\Tilde{y}_{m+1})$ such that
\begin{equation}
\label{eq:free_space_mvt}
    \zeta_m(\mathbf U)
    =
    \left.
    \frac{\partial r(\Tilde{y};\mathbf U)}{\partial \Tilde{y}}
    \right|_{\Tilde{y}=\xi_m}
    g_m .
\end{equation}
Since
\begin{equation}
\label{eq:free_space_slope_bound}
    \left|
    \frac{\partial r(\Tilde{y};\mathbf U)}{\partial \Tilde{y}}
    \right|
    =
    \left|
    \frac{\Tilde{y}-y}
    {c\sqrt{x^2+(y-\Tilde{y})^2+h^2}}
    \right|
    \leq
    \frac{\kappa_D}{c},
\end{equation}
we have
\begin{equation}
\label{eq:free_space_increment_bound}
    |\zeta_m(\mathbf U)|
    \leq
    \frac{g_m \kappa_D}{c}.
\end{equation}
Thus, the guided propagation term provides an exactly linear delay component with respect to the PA spacing, while the free-space propagation term acts as a bounded user-dependent perturbation.

\noindent\textbf{Theorem 1:}
Under the full-aperture constraint $\Tilde{y}_1=0$ and $\Tilde{y}_M=L$, the equal-spacing deployment uniquely minimizes the dispersion of the linear guided-delay increments $\{n_{\mathrm{eff}}g_m/c\}_{m=1}^{M-1}$. Moreover, under the exact spherical-wave delay model, the adjacent-delay dispersion of the equal-spacing deployment is only caused by the bounded free-space perturbation and satisfies
\begin{equation}
\label{eq:uniform_delay_dispersion_bound}
    \mathcal V_{\delta}(\mathbf y^{\mathrm{eq}},\mathbf U)
    \leq
    \frac{\kappa_D^2L^2}{c^2(M-1)},
\end{equation}
where $\mathbf y^{\mathrm{eq}}=[\Tilde{y}_1^{\mathrm{eq}},\ldots,\Tilde{y}_M^{\mathrm{eq}}]^T$ denotes the equal-spacing deployment.

\noindent\textit{Proof:}
Please refer to Appendix A.

\noindent\textbf{Remark 3:}
Theorem~1 shows that the equal-spacing deployment makes the linear guided-delay increments identical for all adjacent PA pairs. Hence, the remaining adjacent-delay dispersion comes only from the bounded free-space perturbation. The bound in \eqref{eq:uniform_delay_dispersion_bound} also reveals the impact of the room geometry. Since $\kappa_D=D_2/\sqrt{D_2^2+h^2}$, a larger ceiling height $h$ reduces the sensitivity of the free-space delay to the PA position, thereby reducing the perturbation bound. A shorter waveguide length $L$ or a larger number of PAs $M$ also decreases this bound because the adjacent PA intervals become shorter. However, this does not mean that arbitrarily reducing $L$ or increasing $M$ is always beneficial. According to \eqref{eq:pair_phase_span} and Lemma~2, a smaller adjacent spacing $g_m=L/(M-1)$ also reduces the tunable phase span of adjacent PA pairs under a finite $\Delta f_{\max}$. Therefore, equal spacing should be interpreted as a robust long-term deployment principle that balances delay regularity and phase-tuning capability, rather than an instantaneous rate-maximizing placement for every user location.

\subsection{Impact of PA-Position-Induced Amplitude Imbalance}
\label{subsec:amplitude_imbalance}

The above deployment analysis is mainly based on the delay structure, since frequency tuning controls the relative phases of the PA-radiated paths. However, PA positions also change the PA-user distances, and hence affect the free-space amplitude coefficients $a_{m,n}=1/d_{m,n}$. Therefore, it is necessary to quantify whether the distance-dependent amplitude imbalance can significantly weaken the phase-alignment gain.
Let
\begin{equation}
\label{eq:average_amplitude}
    \bar a_n
    =
    \frac{1}{M}
    \sum_{m=1}^{M}
    a_{m,n},
\end{equation}
and write
\begin{equation}
\label{eq:epsilon_definition}
    a_{m,n}
    =
    \bar a_n(1+\epsilon_{m,n}),
    \quad
    \sum_{m=1}^{M}
    \epsilon_{m,n}
    =
    0,
\end{equation}
where $\epsilon_{m,n}$ represents the relative amplitude deviation of the $m$-th PA path from the average amplitude. Define the normalized amplitude-imbalance coefficient $\rho_n^2$ as
\begin{equation}
\label{eq:rho_definition}
    \rho_n^2
    =
    \frac{1}{M}
    \sum_{m=1}^{M}
    \epsilon_{m,n}^2,
\end{equation}
which measures the mean squared relative amplitude imbalance among the PA-radiated paths. If all PA paths have identical amplitudes, then $\rho_n^2=0$.

To quantify the gain brought by phase alignment, we compare two reference cases. The first is an incoherent reference, where the PA path phases are not aligned and the pairwise cross terms are averaged out. The second is an ideal coherent reference, where all PA path phases are perfectly aligned. These two references are used only to isolate the effect of amplitude imbalance on the phase-alignment gain.

\noindent\textbf{Lemma 3:}
For a given user $n$, under the amplitude representation in \eqref{eq:epsilon_definition}, the normalized phase-alignment gain from the incoherent reference to the ideal coherent reference is
\begin{equation}
\label{eq:phase_gain}
    \Delta A_{\phi,n}
    =
    M(M-1-\rho_n^2).
\end{equation}
The amplitude-imbalance-induced reduction in the phase-alignment gain is
\begin{equation}
\label{eq:amplitude_imbalance_penalty}
    \Delta A_{\mathrm{amp},n}
    =
    M\rho_n^2.
\end{equation}
Therefore, the relative impact of amplitude imbalance compared with the phase-alignment gain is
\begin{equation}
\label{eq:amplitude_phase_gain_ratio}
    \eta_{\mathrm{amp},n}
    =
    \frac{
    \Delta A_{\mathrm{amp},n}
    }{
    \Delta A_{\phi,n}
    }
    =
    \frac{
    \rho_n^2
    }{
    M-1-\rho_n^2
    }.
\end{equation}

\noindent\textit{Proof:}
Please refer to Appendix B.

For a user located at $\mathbf U$ under a PA deployment scheme $\mathcal S$, let
$\rho^2(\mathbf U,\mathcal S)$ denote the amplitude-imbalance coefficient computed from \eqref{eq:rho_definition}. The corresponding relative impact of PA-position-induced amplitude imbalance is
\begin{equation}
\label{eq:eta_amp_user_deployment}
    \eta_{\mathrm{amp}}(\mathbf U,\mathcal S)
    =
    \frac{
    \rho^2(\mathbf U,\mathcal S)
    }{
    M-1-\rho^2(\mathbf U,\mathcal S)
    } .
\end{equation}
For randomly and uniformly distributed users, the average relative impact under deployment scheme $\mathcal S$ is defined as
\begin{equation}
\label{eq:average_eta_amp_deployment}
    \bar{\eta}_{\mathrm{amp}}^{(\mathcal S)}
    =
    \mathbb E_{\mathbf U,\mathcal S}
    \left[
    \eta_{\mathrm{amp}}(\mathbf U,\mathcal S)
    \right],
\end{equation}
where $\bar{\eta}_{\mathrm{amp}}^{(\mathcal S)}$ is the average relative amplitude-imbalance impact, which measures the PA-position-induced free-space amplitude imbalance relative to the phase-alignment gain. The expectation is taken over user locations and, when applicable, over the random PA deployment realizations. For deterministic deployments, the expectation is only over $\mathbf U$.

The quantity $\bar{\eta}_{\mathrm{amp}}^{(\mathcal S)}$ is not used to measure the absolute path-loss variation. Instead, it measures the average impact of PA-position-induced amplitude imbalance relative to the phase-alignment gain in \eqref{eq:phase_gain}. Hence, when $\bar{\eta}_{\mathrm{amp}}^{(\mathcal S)}\ll1$, the free-space amplitude variation among different PA-assisted paths is much weaker than the coherent-combining gain that can be recovered by phase alignment.

\noindent\textbf{Remark 4:}
As will be shown by the Monte Carlo results in Section~\ref{sec:numerical_results}, $\bar{\eta}_{\mathrm{amp}}^{(\mathcal S)}$ remains small under the considered indoor setup. This supports the phase-oriented design principle adopted in this paper: PA deployment and frequency tuning are mainly guided by the controllable phase structure, while the distance-dependent amplitude imbalance caused by PA positions is treated as a secondary perturbation. Nevertheless, the amplitude effect is not discarded in performance evaluation, since the final SNR and achievable rate are computed using the exact weighted factor $A_{w,n}(\Delta f_n)$.

\subsection{Approximate Closed-Form Offset Under Equal Spacing}
\label{subsec:approximate_closed_form_offset}

The equal-spacing deployment makes the relative delays with respect to a reference PA approximately form an arithmetic progression, which can be used to derive a closed-form approximate frequency offset for the multi-PA case.

For user $n$, define the average adjacent relative delay as
\begin{equation}
\label{eq:average_relative_delay_equal}
    \bar\delta_n
    =
    \frac{\tau_{M,n}-\tau_{1,n}}{M-1}.
\end{equation}
Under the arithmetic-delay approximation, the relative delay with respect to PA 1 is approximated by
\begin{equation}
\label{eq:equal_relative_delay_approx}
    \tau_{m,n}-\tau_{1,n}
    \approx
    (m-1)\bar\delta_n,
    \quad
    m=2,\ldots,M.
\end{equation}
According to Lemma~3, the PA-position-induced amplitude imbalance acts as a secondary perturbation compared with the phase-alignment gain. Therefore, to obtain a tractable closed-form approximation, we focus on the dominant phase-only coherent-combining factor. Removing the common phase of PA 1 and substituting \eqref{eq:equal_relative_delay_approx} gives
\begin{equation}
\label{eq:dirichlet_array_factor}
    A_n^{\mathrm{ari}}(\Delta f_n)
    =
    \left|
    \sum_{q=0}^{M-1}
    e^{-j2\pi(f_c+\Delta f_n)q\bar\delta_n}
    \right|^2.
\end{equation}
The summation in \eqref{eq:dirichlet_array_factor} is a finite geometric series. Hence, it can be equivalently written in the Dirichlet-kernel form as
\begin{equation}
\label{eq:dirichlet_closed_form}
    A_n^{\mathrm{ari}}(\Delta f_n)
    =
    \left|
    \frac{
    \sin\left(
    \pi M(f_c+\Delta f_n)\bar\delta_n
    \right)
    }{
    \sin\left(
    \pi(f_c+\Delta f_n)\bar\delta_n
    \right)
    }
    \right|^2.
\end{equation}

\noindent\textbf{Proposition 4:}
For the approximate Dirichlet-kernel factor $A_n^{\mathrm{ari}}(\Delta f_n)$ in \eqref{eq:dirichlet_closed_form}, the maximum value $M^2$ is achieved if and only if
\begin{equation}
\label{eq:ari_integer_condition}
    (f_c+\Delta f_n)\bar\delta_n \in \mathbb Z.
\end{equation}
Let $\mathcal I_n^{\mathrm{ari}}=[(f_c-\Delta f_{\max})\bar\delta_n,(f_c+\Delta f_{\max})\bar\delta_n]$ denote the feasible normalized phase-cycle interval. If $\mathbb Z\cap\mathcal I_n^{\mathrm{ari}}\neq\emptyset$, the approximate closed-form offset
\begin{equation}
\label{eq:ari_closed_form_offset_general}
    \Delta f_{n,\mathrm{ari}}
    =
    \frac{k_n^\star}{\bar\delta_n}-f_c,
    \quad
    k_n^\star\in\mathbb Z\cap\mathcal I_n^{\mathrm{ari}},
\end{equation}
attains the maximum of $A_n^{\mathrm{ari}}(\Delta f_n)$ over the feasible interval.

\noindent\textit{Proof:}
Let $x=(f_c+\Delta f_n)\bar\delta_n$. Then $A_n^{\mathrm{ari}}=\big|\sum_{q=0}^{M-1}e^{-j2\pi qx}\big|^2\leq M^2$ by the triangle inequality, with equality if and only if all summands share the same phase, i.e., $x\in\mathbb Z$. If an integer $k_n^\star\in\mathcal I_n^{\mathrm{ari}}$ exists, the offset in \eqref{eq:ari_closed_form_offset_general} is feasible and attains this maximum. This completes the proof.

\noindent\textbf{Remark 5 (Infeasible case):}
If $\mathbb Z\cap\mathcal I_n^{\mathrm{ari}}=\emptyset$, we adopt the nearest-integer endpoint rule
\begin{equation}
    \Delta f_{n,\mathrm{ari}}
    =
    \frac{x_n^\star}{\bar\delta_n}-f_c,
    \quad
    x_n^\star=\operatorname*{arg\,min}_{x\in\mathcal I_n^{\mathrm{ari}}}\operatorname{dist}(x,\mathbb Z),
\end{equation}
as a low-complexity heuristic. Unlike the two-PA case, this rule is not guaranteed to maximize $A_n^{\mathrm{ari}}(\Delta f_n)$ over an integer-free interval, since the Dirichlet kernel exhibits sidelobes between adjacent mainlobes; the resulting offset can be refined by Algorithm~\ref{alg:multi_pa_frequency} if needed.

The closed-form offset in Proposition~4 is derived from the arithmetic relative-delay approximation under equal-spacing deployment. Its accuracy depends on how closely the actual relative delays follow \eqref{eq:equal_relative_delay_approx}. When the free-space perturbation is small, it provides a low-complexity frequency-offset solution; otherwise, it can serve as an initialization or benchmark for Algorithm~\ref{alg:multi_pa_frequency}.

\section{Numerical Results}
\label{sec:numerical_results}

In this section, numerical results are presented to evaluate the performance of the proposed frequency-tuned beamforming design for PASS. The simulations examine the frequency-tuning behavior, PA-deployment effect, worst-user performance, and PA-scaling trend, thereby illustrating the analytical insights developed in the previous sections.

Unless otherwise specified, the room size is $5~\mathrm{m}\times5~\mathrm{m}\times3~\mathrm{m}$, the effective refractive index is $n_{\mathrm{eff}}=1.4$, the transmit power is $P=-20$ dBm, and the bandwidth is $10$ MHz. The AWGN power is computed as $-174+10\log_{10}(\mathrm{BW})$ dBm, where $\mathrm{BW}$ is the bandwidth in Hz. Users are randomly and uniformly distributed over the floor region. Accordingly, the user coordinates are modeled as independent random variables with $x_n\sim\mathcal{U}[0,D_1]$ and $y_n\sim\mathcal{U}[0,D_2]$. The one-dimensional search in Algorithm~\ref{alg:multi_pa_frequency} uses $Q=2000$ grid points, and each Monte Carlo curve is averaged over $10^6$ random samples.

\subsection{Frequency-Offset Coherent-Combining Capability}

\begin{figure}[t]
\centering
\includegraphics[width=0.9\linewidth]{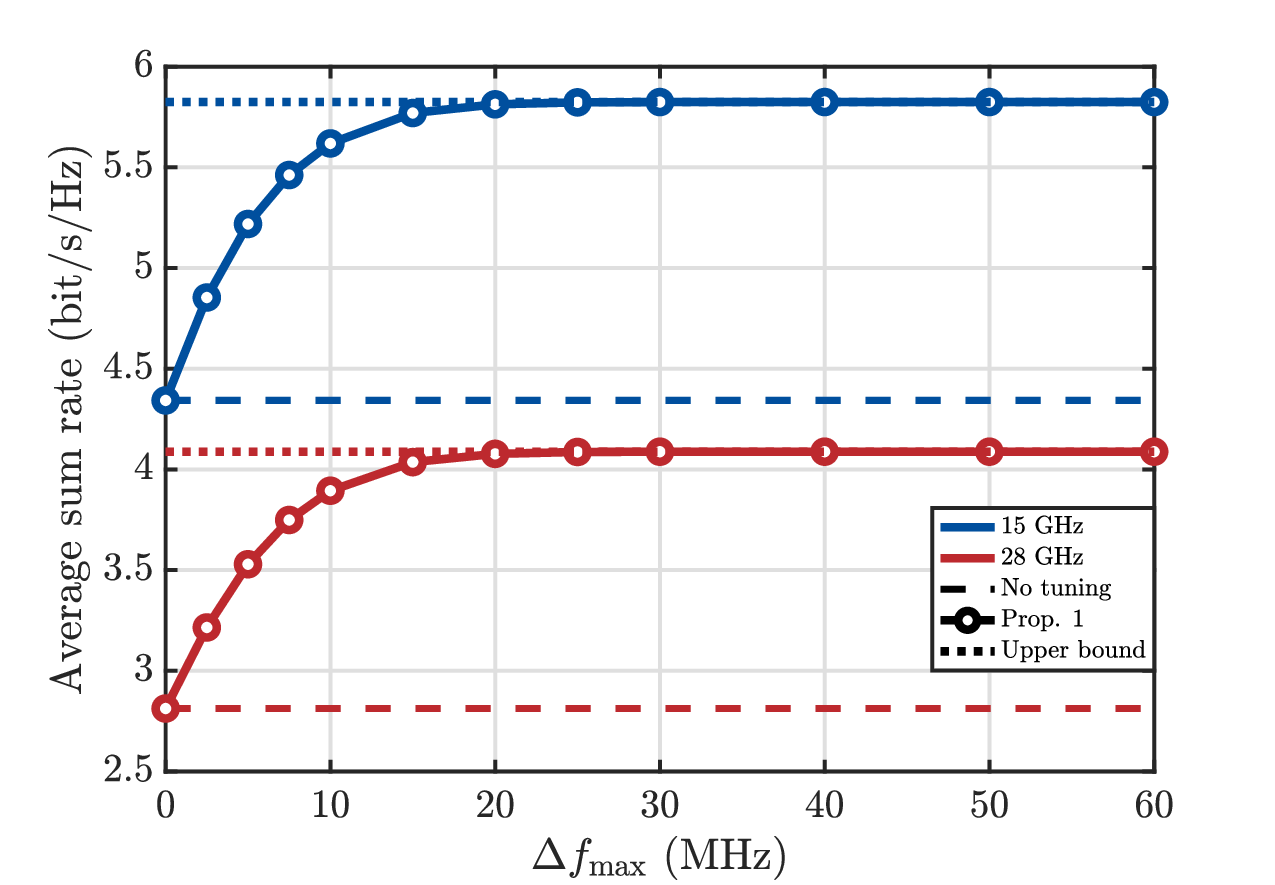}
\caption{Average sum rate versus $\Delta f_{\max}$ for the two-PA configuration under the full-aperture equal-spacing deployment.}
\label{fig:dfmax_m2_fc}
\end{figure}

\begin{figure}[t]
\centering
\includegraphics[width=0.9\linewidth]{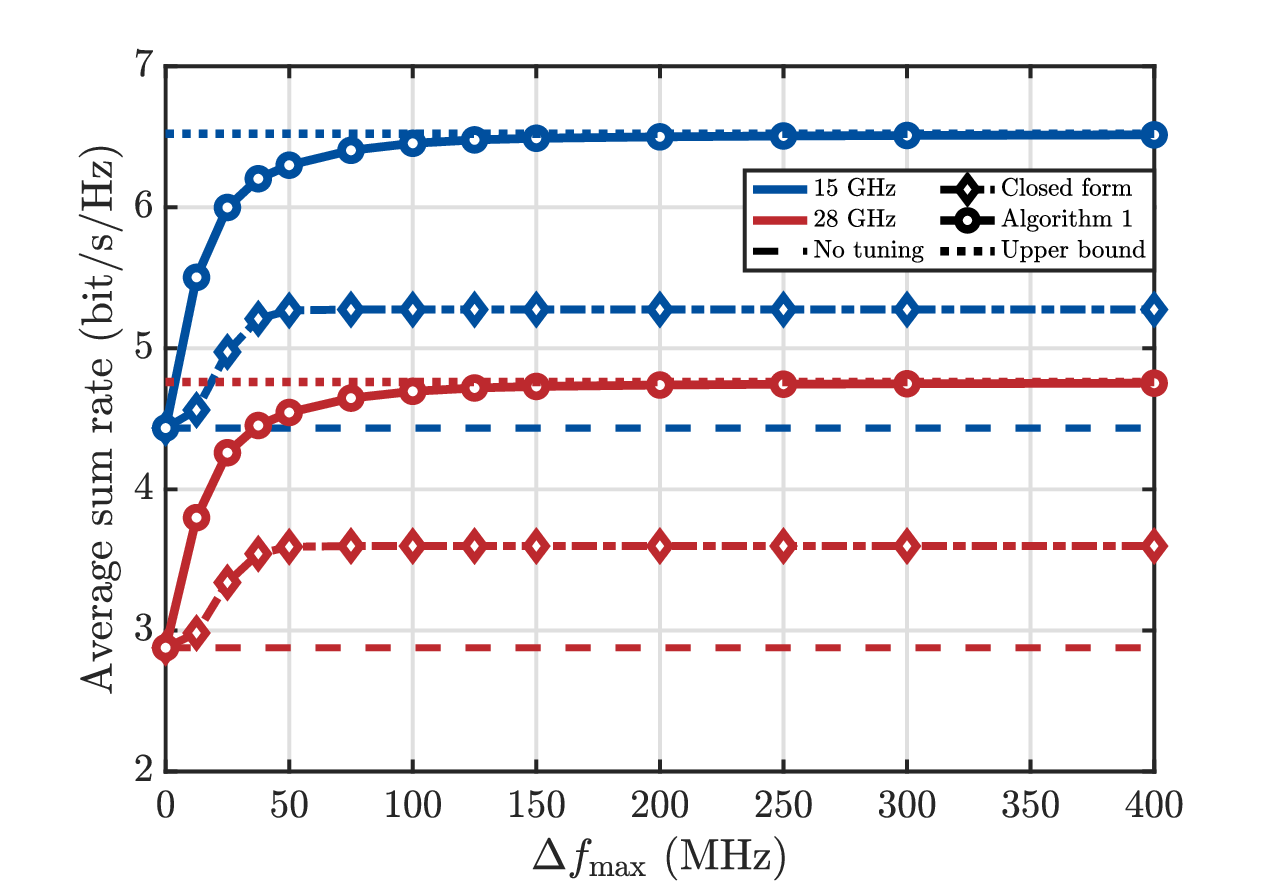}
\caption{Average sum rate versus $\Delta f_{\max}$ for the three-PA configuration under the full-aperture equal-spacing deployment.}
\label{fig:dfmax_m3_fc}
\end{figure}

Fig.~\ref{fig:dfmax_m2_fc} shows the average sum rate of the two-PA configuration at $15$ GHz and $28$ GHz. The no-tuning curve corresponds to the fixed-frequency baseline with $\Delta f_n=0$, while the coherent upper bound represents the ideal case where all PA-assisted signals are perfectly phase-aligned at the receiver. Since the two-PA case contains only one relative phase term, Proposition~1 provides a closed-form optimal frequency offset. As $\Delta f_{\max}$ increases, the proposed closed-form design rapidly improves the achievable rate and approaches the coherent upper bound once the feasible interval includes a near-coherent phase point. After this point, further increasing $\Delta f_{\max}$ brings little additional gain. The result at $28$ GHz follows the same trend as that at $15$ GHz, but its rate is lower because the free-space path loss is larger at the higher carrier frequency.

Fig.~\ref{fig:dfmax_m3_fc} presents the corresponding three-PA result. The closed-form approximation in Proposition~4 improves over no tuning and approaches its saturated performance at a relatively small $\Delta f_{\max}$. This indicates that the closed-form offset can provide a useful low-complexity candidate, although its maximum achievable performance over the considered frequency-offset range remains below that of Algorithm~\ref{alg:multi_pa_frequency}. The performance gap is due to the fact that Proposition~4 is based on an approximate arithmetic-delay representation, whereas the actual PA-user relative delays are affected by the exact spherical-wave propagation distances. Algorithm~\ref{alg:multi_pa_frequency}, in contrast, evaluates the exact weighted coherent-combining objective at all feasible stationary-point candidates and interval endpoints, and thus approaches the coherent upper bound more closely.

Figs.~\ref{fig:dfmax_m2_fc} and \ref{fig:dfmax_m3_fc} also support the observation in Remark~1. For the same PA geometry, the additional phase excursion induced by frequency tuning is governed by the product of the frequency offset and the PA-user propagation-delay difference, rather than by the absolute carrier frequency $f_c$.
Therefore, the curves at $15$ GHz and $28$ GHz reach their plateaus at nearly the same maximum frequency offset.
The carrier frequency mainly affects the initial phase state and the path-loss level, so the $28$ GHz curves exhibit lower absolute rates than the $15$ GHz curves due to the larger free-space path loss while requiring a similar frequency-offset range to recover most of the coherent-combining gain.

\subsection{Deployment-Aware Performance, Fairness, and PA Scaling}

\begin{figure*}[!t]
\centering
\subfloat[$M=2$.]{
\includegraphics[width=0.45\textwidth]{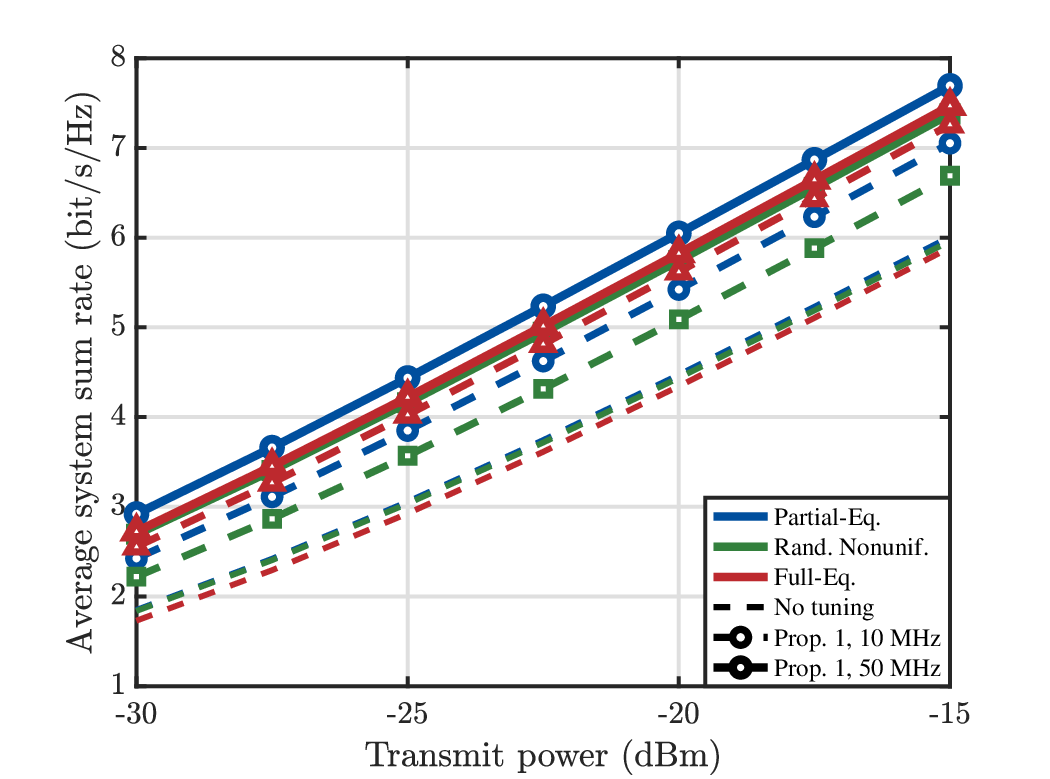}
\label{fig:avg_sum_power_m2}}
\hfill
\subfloat[$M=3$.]{
\includegraphics[width=0.45\textwidth]{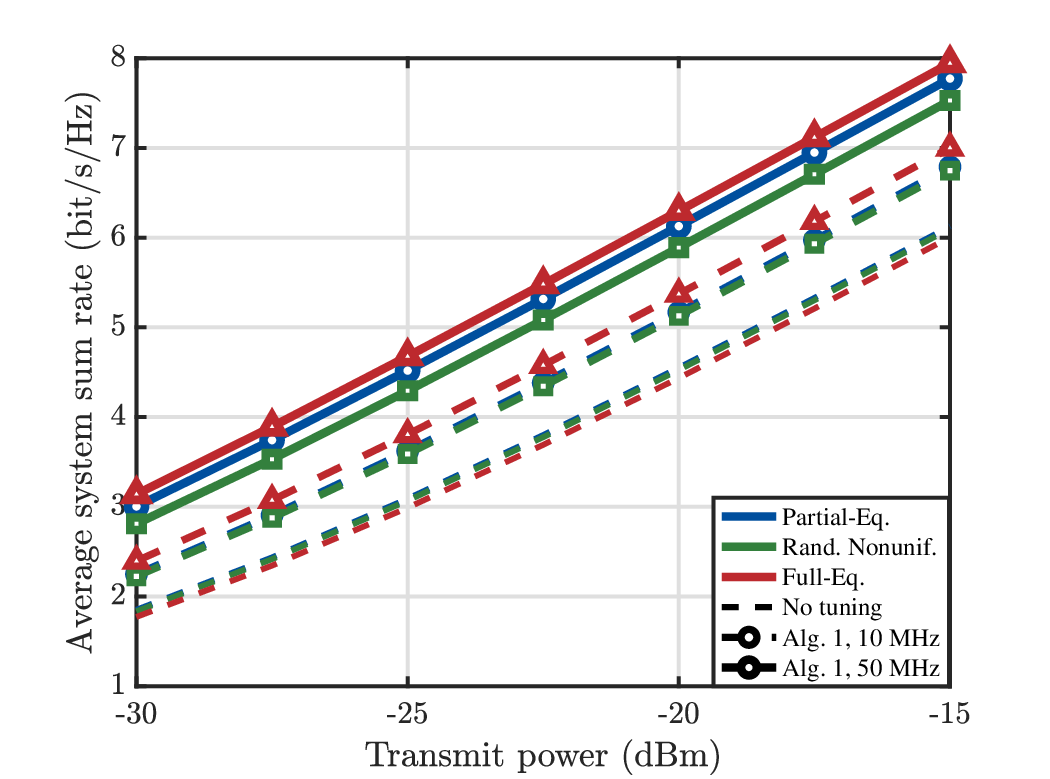}
\label{fig:avg_sum_power_m3}}
\caption{Average system sum rate versus transmit power under different PA deployments at $f_c=15$ GHz.}
\label{fig:avg_sum_power_deployment}
\end{figure*}

\begin{figure*}[!t]
\centering
\subfloat[$N=4$, $M=3$.]{
\includegraphics[width=0.33\textwidth]{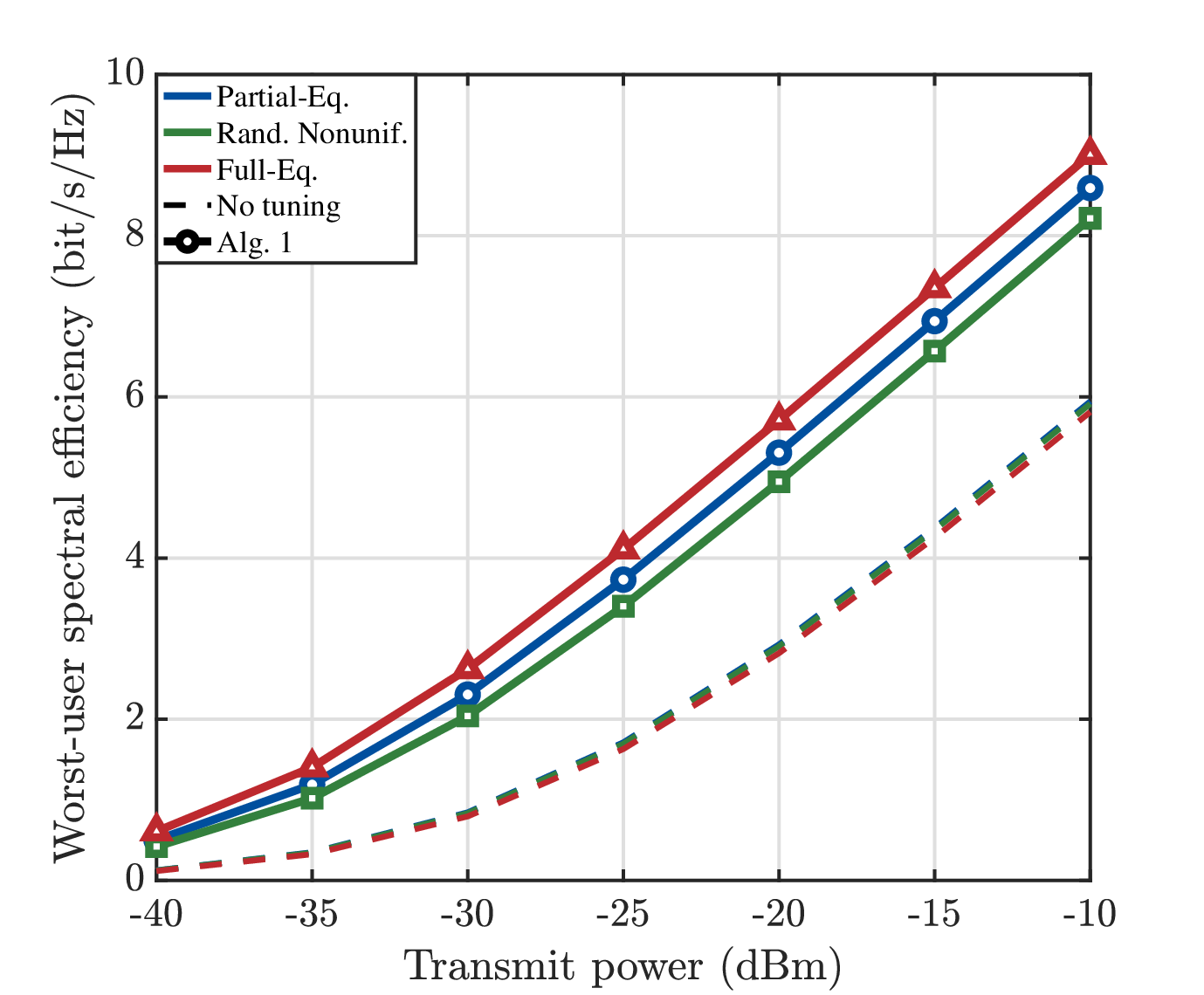}
\label{fig:worst_user_n4}}
\subfloat[$N=8$, $M=3$.]{
\includegraphics[width=0.33\textwidth]{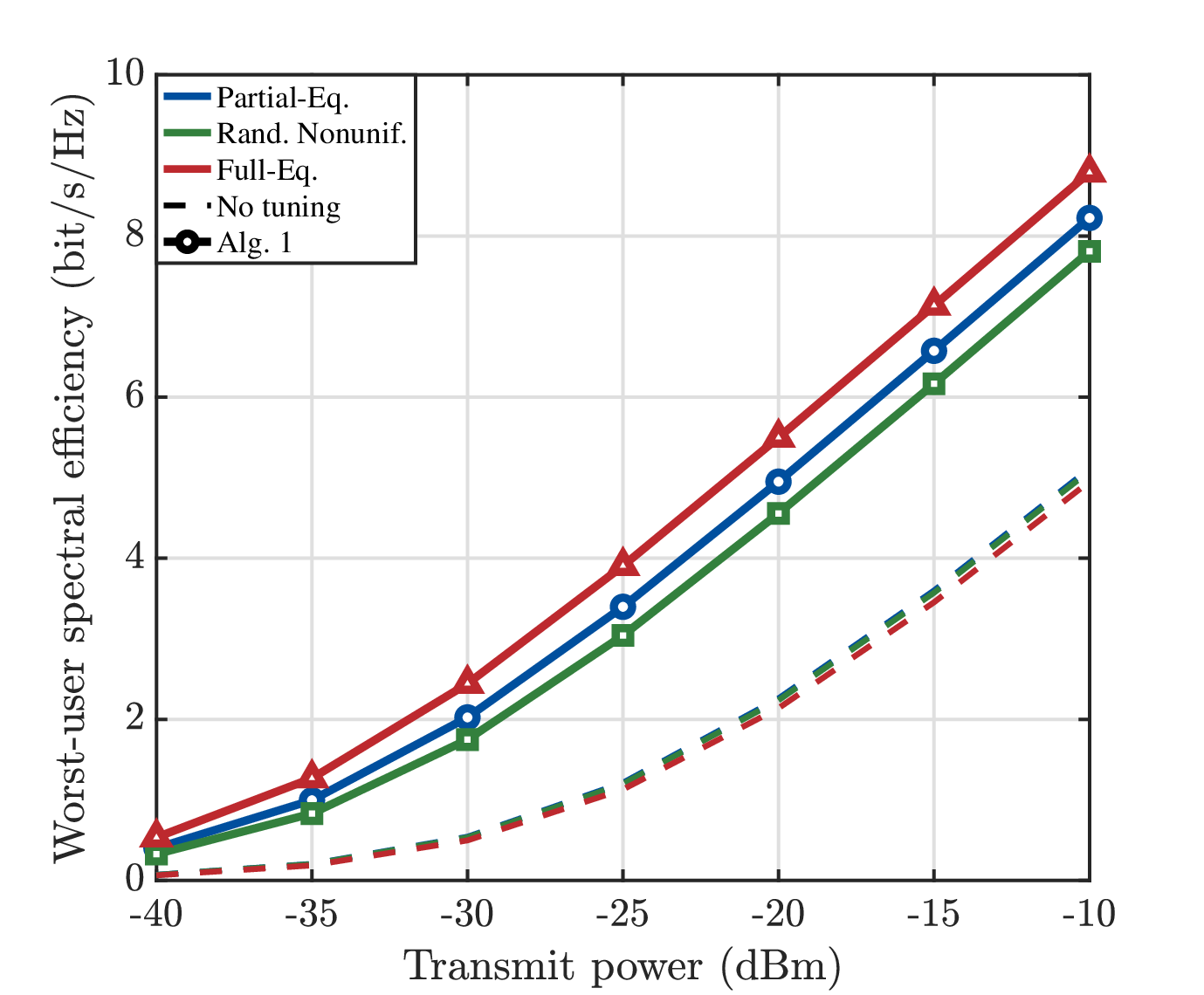}
\label{fig:worst_user_n8}}
\subfloat[$N=16$, $M=3$.]{
\includegraphics[width=0.33\textwidth]{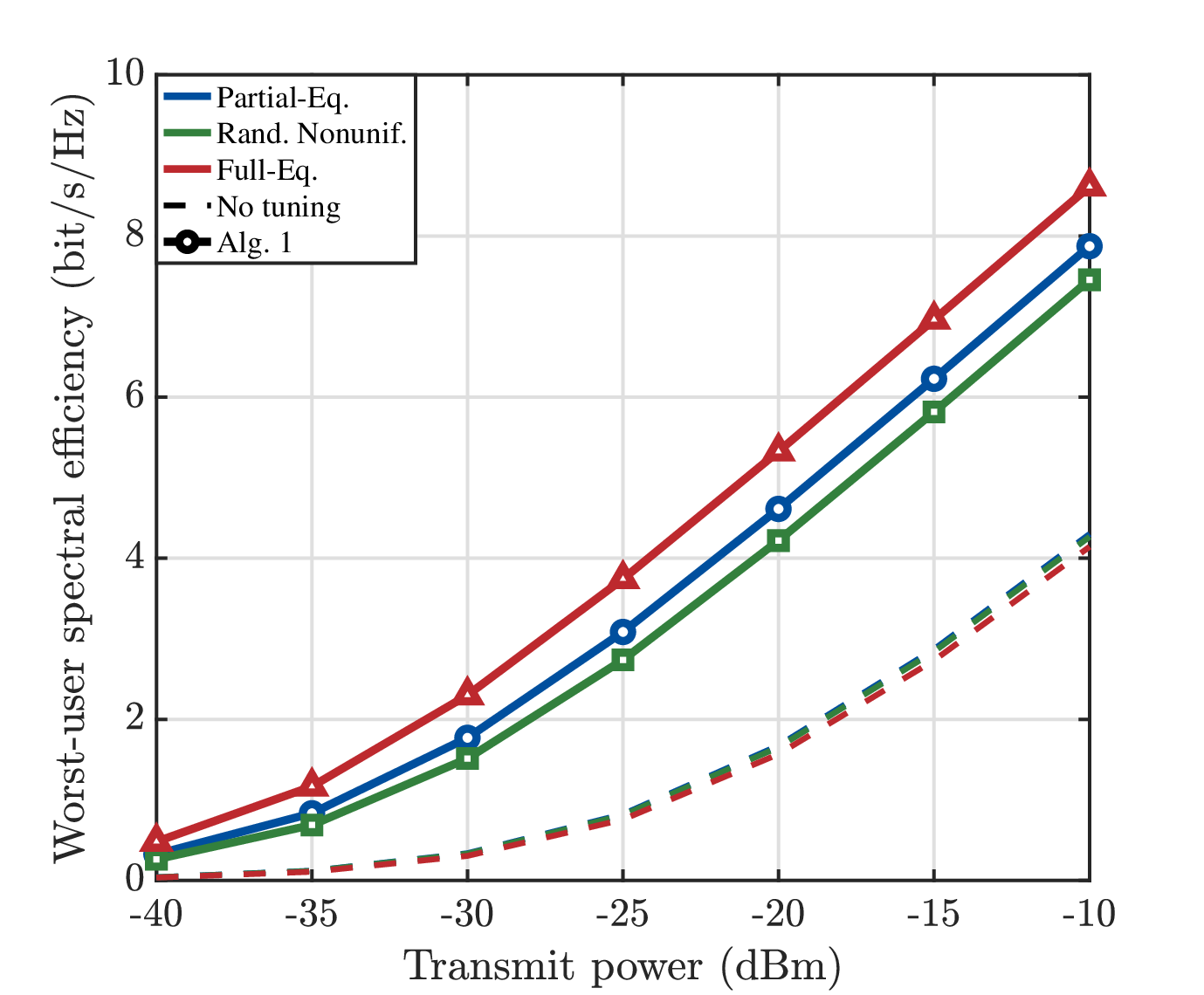}
\label{fig:worst_user_n16}}
\caption{Average worst-user spectral efficiency (without the TDMA pre-log factor) versus transmit power under different PA deployments for $M=3$, $\Delta f_{\max}=50$ MHz, and $f_c=15$ GHz.}
\label{fig:worst_user_deployment}
\end{figure*}

We next compare three PA deployment schemes along the ceiling-edge waveguide. The full-aperture equal deployment places the PAs as $\Tilde{y}_m=(m-1)D_2/(M-1)$, which follows the feasible deployment strategy suggested by the PA-deployment analysis in Section~\ref{sec:pa_deployment_frequency_tuned}. This deployment uses the entire available aperture and yields a regular PA spacing that is favorable for single-offset phase alignment. The partial-aperture equal deployment places equally spaced PAs over half of the full aperture, where the starting point of the half-aperture segment is drawn uniformly for each Monte Carlo realization. The random nonuniform deployment draws ordered PA locations over the full aperture subject only to the minimum spacing constraint $\Tilde{y}_{m+1}-\Tilde{y}_m\geq\Delta_{\min}$. Hence, both the partial-aperture equal and random nonuniform deployments vary across Monte Carlo realizations, while the full-aperture equal deployment is deterministic.

Fig.~\ref{fig:avg_sum_power_deployment} compares the average system sum rate for $\Delta f_{\max}=10$ MHz and $50$ MHz. Frequency tuning improves all deployments over the fixed-frequency baseline, and a larger tuning range yields a higher rate owing to the wider feasible interval for phase alignment. For $M=2$, the partial-aperture equal deployment slightly outperforms the full-aperture one when $\Delta f_{\max}=50$ MHz: the half-aperture spacing already provides sufficient delay difference for phase alignment, while the PAs located closer to the room center reduce the distance-dependent path loss for many user realizations. When $\Delta f_{\max}=10$ MHz, however, the shorter spacing limits the tunable phase excursion, and the partial-aperture deployment becomes inferior. For $M=3$, the full-aperture equal deployment is favorable under both budgets, since adding the third PA reduces the adjacent spacing and introduces more PA pairs that must be jointly aligned by a single offset, in which case the larger delay spread offered by the full aperture becomes essential for creating sufficient tunable phase excursion. Moreover, both equal-spacing deployments outperform the random nonuniform deployment, which confirms the benefit of regular PA spacing for multi-PA frequency-tuned beamforming.

Fig.~\ref{fig:worst_user_deployment} evaluates the average worst-user spectral efficiency before TDMA scaling for $N=4$, $8$, and $16$ users under $\Delta f_{\max}=50$ MHz, where the $1/N$ pre-log factor is removed to enable a clearer comparison across different user numbers.
In each user drop, all users share the same three-PA deployment, and a new deployment is independently generated for the next drop. For the no-tuning baseline, the worst-user spectral efficiency decreases noticeably as the user number increases, because a larger user set is more likely to contain a user located near a destructive-combining region. In contrast, Algorithm~\ref{alg:multi_pa_frequency} maintains a much more stable worst-user performance across different user numbers, showing that frequency tuning can effectively move disadvantaged users away from destructive phase conditions. Compared with the average-rate results, the gain in the worst-user performance is more pronounced. This is because frequency tuning brings limited additional benefit when a user is already close to a favorable phase-alignment condition, whereas it can provide a substantial improvement when the PA-assisted signals are nearly out of phase. Therefore, frequency-tuned phase alignment effectively mitigates destructive combining and improves the worst-case robustness of PASS. 
The full-aperture equal deployment achieves the strongest worst-user performance in the considered setting, which agrees with the deployment analysis: regular spacing avoids poorly controllable adjacent PA pairs and provides a delay structure that is more compatible with single-offset phase alignment.

\begin{figure}[t]
\centering
\includegraphics[width=0.9\linewidth]{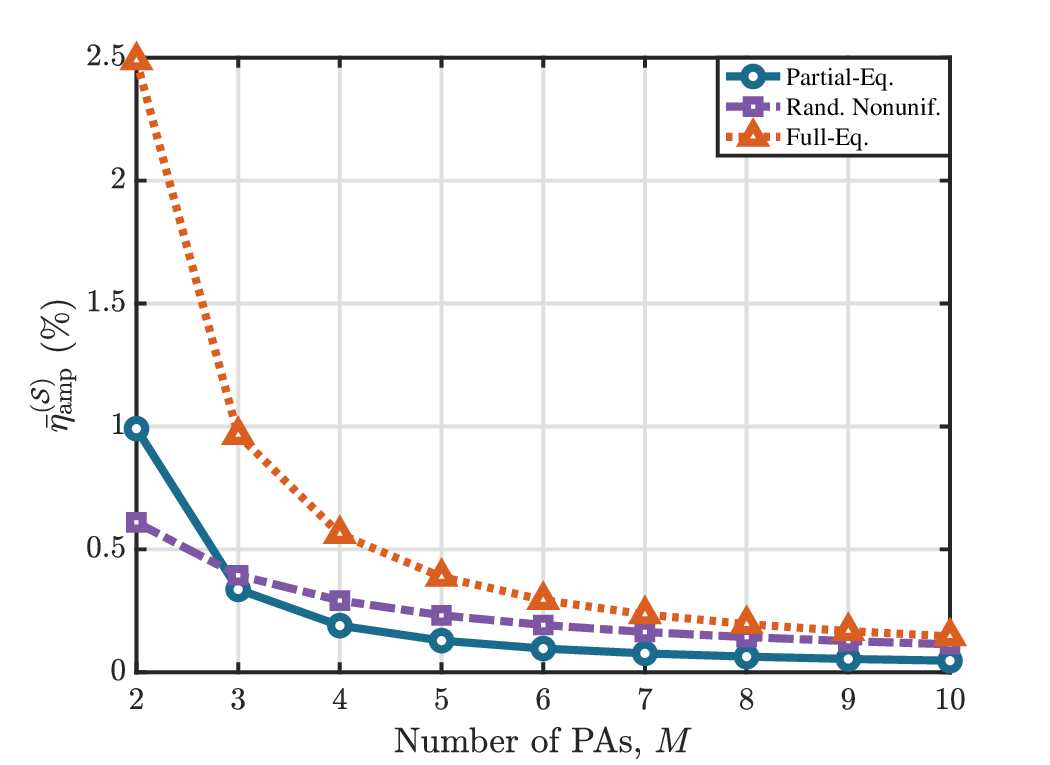}
\caption{Average relative amplitude-imbalance impact $\bar{\eta}_{\mathrm{amp}}^{(\mathcal S)}$ versus the number of PAs under different PA deployments.}
\label{fig:amplitude_imbalance}
\end{figure}

Fig.~\ref{fig:amplitude_imbalance} plots the average impact of PA-position-induced free-space amplitude imbalance relative to the phase-alignment gain $\bar{\eta}_{\mathrm{amp}}^{(\mathcal S)}$ for different PA deployment schemes. For all three deployment schemes, $\bar{\eta}_{\mathrm{amp}}^{(\mathcal S)}$ decreases as $M$ increases. This is because more PAs average the distance-dependent amplitude variations over more PA-assisted paths, while the phase-alignment gain increases with the number of PA pairs. The full-aperture equal deployment has the largest value, since spreading the PAs over the whole aperture increases the possible distance difference between different PA-user paths. Nevertheless, even for the two-PA case, the relative impact of free-space amplitude imbalance is only about $2.5\%$ of the phase-alignment gain in the considered setup. Moreover, although the full-aperture equal deployment suffers the strongest amplitude-imbalance effect among the three schemes, the rate results in Figs.~\ref{fig:avg_sum_power_deployment} and \ref{fig:worst_user_deployment} show that it still achieves the best performance. This further confirms that phase alignment, rather than distance-dependent amplitude variation, is the dominant factor in the considered multi-PA PASS design.

\begin{figure}[t]
\centering
\includegraphics[width=0.9\linewidth]{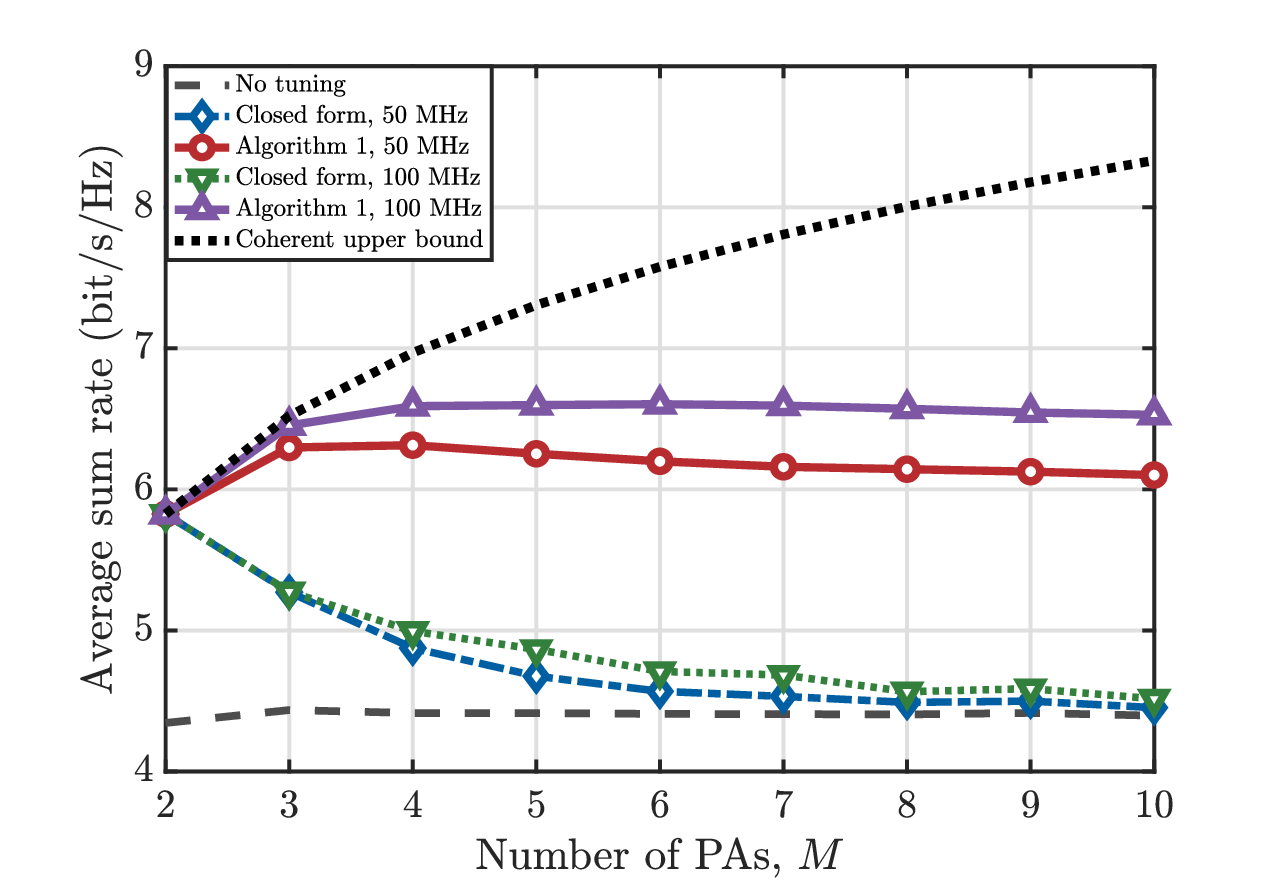}
\caption{Average sum rate versus the number of PAs under the full-aperture equal deployment at $f_c=15$ GHz.}
\label{fig:rate_vs_m}
\end{figure}

Fig.~\ref{fig:rate_vs_m} shows the average sum rate versus the PA number under the full-aperture equal deployment for $\Delta f_{\max}=50$ MHz and $100$ MHz.
The no-tuning baseline changes only mildly as the PA number increases, showing that simply adding more PAs does not necessarily bring a rate gain if the PA-assisted signals are not phase-aligned. The coherent upper bound increases with the PA number, since more PA-assisted paths can contribute constructively when ideal phase alignment is achieved. 
Algorithm~\ref{alg:multi_pa_frequency} achieves a clearly higher rate than the no-tuning baseline, demonstrating the benefit of frequency-tuned phase alignment. However, as the PA number increases, its gap to the coherent upper bound gradually becomes larger. This is because, under a fixed aperture, adding more PAs reduces the adjacent PA spacing and limits the tunable phase excursion between neighboring PA-assisted paths; meanwhile, more pairwise phase relations need to be jointly controlled by one scalar frequency offset. The closed-form design also approaches the no-tuning baseline as the PA number increases, indicating that the equal-delay approximation is more effective for a small number of PAs and becomes less accurate when the multi-PA phase relations become more complicated.

\section{Conclusion}
In this paper, we proposed FreqTune-PASS, a frequency-tuned beamforming framework for multi-user downlink PASS, where all PA positions remain fixed across TDMA slots and only a user-specific carrier-frequency offset is adapted for the scheduled user. Based on a delay-domain model, we characterized the pairwise coherent-combining mechanism of frequency tuning, derived a closed-form globally optimal offset for the two-PA case, and developed a low-complexity one-dimensional search together with an approximate closed-form offset for the multi-PA case. We further proved that the full-aperture equal-spacing deployment maximizes the guaranteed phase-tuning span and renders the relative delays approximately an arithmetic progression for uniformly distributed users. Numerical results demonstrated that the proposed design substantially improves both the sum rate and the worst-user rate over fixed-frequency transmission. 
Moreover, the rate achieved by frequency tuning approaches the coherent-combining upper bound, which establishes frequency tuning as a practical alternative to PA repositioning.

\appendices

\section{Proof of Theorem 1}
From \eqref{eq:delta_guided_residual}, the linear guided-delay increment between adjacent PAs is $(n_{\mathrm{eff}}/c)g_m$, whose dispersion is
\begin{equation}
\label{eq:guided_delay_nonuniformity}
    \mathcal V_{\mathrm{gd}}(\mathbf y)=\sum_{m=1}^{M-1}\Big(\frac{n_{\mathrm{eff}}}{c}g_m-\frac{n_{\mathrm{eff}}}{c}\bar g\Big)^2,
\end{equation}
where $\bar g=\frac{1}{M-1}\sum_{m=1}^{M-1}g_m$ is the average adjacent spacing. Since $n_{\mathrm{eff}}/c$ is a common factor, minimizing $\mathcal V_{\mathrm{gd}}(\mathbf y)$ is equivalent to minimizing $\sum_{m=1}^{M-1}(g_m-\bar g)^2$. 

Under the full-aperture constraint, $\sum_{m=1}^{M-1}g_m=L$ and thus $\bar g=L/(M-1)$. Hence, minimizing the guided-delay dispersion is equivalent to minimizing $\sum_{m=1}^{M-1}(g_m-L/(M-1))^2$. This strictly convex quadratic function is uniquely minimized when $g_1=\cdots=g_{M-1}=L/(M-1)$, which gives the equal-spacing deployment in \eqref{eq:uniform_pa_deployment}.

For the equal-spacing deployment, the guided component is identical for all adjacent PA pairs, so the adjacent-delay dispersion is only due to the perturbation terms $\{\zeta_m(\mathbf U)\}$. With $\bar\zeta(\mathbf U)=\frac{1}{M-1}\sum_{m=1}^{M-1}\zeta_m(\mathbf U)$,
\begin{equation}
    \mathcal V_{\delta}(\mathbf y^{\mathrm{eq}},\mathbf U)=\sum_{m=1}^{M-1}\big(\zeta_m(\mathbf U)-\bar\zeta(\mathbf U)\big)^2
    \leq\sum_{m=1}^{M-1}\zeta_m^2(\mathbf U).
\end{equation}
Using $|\zeta_m(\mathbf U)|\leq g_m\kappa_D/c$ and $g_m=L/(M-1)$ gives
\begin{equation}
    \mathcal V_{\delta}(\mathbf y^{\mathrm{eq}},\mathbf U)\leq(M-1)\Big(\frac{\kappa_D L}{c(M-1)}\Big)^2=\frac{\kappa_D^2L^2}{c^2(M-1)},
\end{equation}
which completes the proof.

\section{Proof of Lemma 3}
Using $a_{m,n}=\bar a_n(1+\epsilon_{m,n})$, the normalized weighted coherent-combining factor can be written as
\begin{equation}
    \frac{A_{w,n}}{\bar a_n^2}=\Big|\sum_{m=1}^{M}(1+\epsilon_{m,n})e^{-j\theta_{m,n}}\Big|^2,
\end{equation}
where $\theta_{m,n}$ denotes the phase of the path through PA $m$. Under the random-phase reference model, $\mathbb E_{\phi}[e^{-j(\theta_{m,n}-\theta_{\ell,n})}]=0$ for $m\neq\ell$, so the pairwise cross terms vanish on average and
\begin{equation}
    \mathbb E_{\phi}\Big[\frac{A_{w,n}^{\mathrm{incoh}}}{\bar a_n^2}\Big]
    =\sum_{m=1}^{M}(1+\epsilon_{m,n})^2
    =M(1+\rho_n^2),
\end{equation}
where $A_{w,n}^{\mathrm{incoh}}$ denotes the incoherent reference and the last equality follows from $\sum_{m}\epsilon_{m,n}=0$ and \eqref{eq:rho_definition}. For the ideal coherent reference $A_{w,n}^{\mathrm{coh}}$, all path phases are aligned, and $\sum_{m}\epsilon_{m,n}=0$ yields
\begin{equation}
    \frac{A_{w,n}^{\mathrm{coh}}}{\bar a_n^2}=\Big(\sum_{m=1}^{M}(1+\epsilon_{m,n})\Big)^2=M^2.
\end{equation}

Thus, the normalized phase-alignment gain is $\Delta A_{\phi,n}=M^2-M(1+\rho_n^2)=M(M-1-\rho_n^2)$.  Compared with the imbalance-free gain $M(M-1)$, the reduction caused by amplitude imbalance is $\Delta A_{\mathrm{amp},n}=M\rho_n^2$. The ratio of this reduction to the actual phase-alignment gain is
\begin{equation}
    \eta_{\mathrm{amp},n}
    =
    \frac{M\rho_n^2}{M(M-1-\rho_n^2)}
    =
    \frac{\rho_n^2}{M-1-\rho_n^2},
\end{equation}
which completes the proof.

\bibliographystyle{IEEEtran}
\bibliography{FreqTuned_PASS}


\vspace{11pt}

\vfill

\end{document}